\documentclass[10pt,onecolumn]{IEEEtran}

\def\id{{\hbox{\Bbb I}}}

\def\duzomniejsze{<\kern-.7mm<}
\def\duzowieksze{>\kern-.7mm>}

\def\textbf#1{{\bf #1}}
\def\beq{\begin{equation}}
\def\eeq{\end{equation}}
\def\be{\begin{equation}}
\def\ee{\end{equation}}
\def\ben{\begin{eqnarray}}
\def\een{\end{eqnarray}}
\def\beqa{\begin{eqnarray}}
\def\eeqa{\end{eqnarray}}
\def\eea{\end{array}}
\def\bea{\begin{array}}
\newcommand{\bei}{\begin{itemize}}
\newcommand{\eei}{\end{itemize}}
\newcommand{\bee}{\begin{enumerate}}
\newcommand{\eee}{\end{enumerate}}

\def \s {\,\,\,\,}

\def\rab{\rho_{AB}}

\def\hcal{{\cal H}}
\def\scal{{\cal S}}
\def\pcal{{\cal P}}
\def\bcal{{\cal B}}
\def\ccal{{\cal C}}
\def\wcal{{\cal W}}
\def\ucal{{\cal U}}

\def\>{\rangle}
\def\<{\langle}
\def\tr{{\rm Tr}}
\def\blacksquare{\vrule height 4pt width 4pt depth0pt}
\def\rh{\rho_{ABA'B'}}
\def\ep{\epsilon}
\def\saabb{\sigma_{ABA'B'}}

\def\rhoreal{\rho_{real}^{ccq}}
\def\rhoideal{\rho_{ideal}^{ccq}}

\def\pdit{pdit}
\def\gamt{\gamma^{(2)}_{ABA'B'}}
\def\pdits{pdits}
\def\pbit{pbit}
\def\gamd{\gamma^{(d)}_{ABA'B'}}
\def\pbits{pbits}
\def\Pbits{Pbits}
\def\ccq{ccq }
\def\kd{K_D}

\def\id{I}
\def\main{key part}
\def\mains{key parts}
\def\side{shield}
\def\sides{shields}
\def\cohsq{privacy squeezing}
\def\cosq{p-squeezing}
\def\cosqd{p-squeezed}

\def\ot{\otimes}
\newtheorem{lemma}{Lemma}
\newtheorem{corrolary}{Corrolary}
\newtheorem{theorem}{Theorem}
\newtheorem{defin}{Definition}
\newtheorem{rem}{Remark}

\newtheorem{prop}{Proposition}

\begin{document}

\title{General paradigm for distilling classical key from quantum states}
\author{Karol Horodecki,\thanks{Karol Horodecki, Department of Mathematics Physics and Computer Science, 
University of Gda\'nsk,  Gda\'nsk, Poland}
 Micha{\l} Horodecki,\thanks{Micha{\l} Horodecki, Institute of Theoretical Physics and Astrophysics, 
University of Gda\'nsk, Gda\'nsk, Poland}
 Pawe{\l} Horodecki,\thanks{Pawe{\l} Horodecki, Faculty of Applied Physics and Mathematics, 
 Technical University of Gda\'nsk, Gda\'nsk, Poland}
 Jonathan Oppenheim\thanks{Jonathan Oppenheim, Department of Applied Mathematics and Theoretical Physics, 
University of Cambridge U.K.}
 }
\maketitle
\author{ Karol Horodecki$^{(1)}$, Micha\l{} Horodecki$^{(2)}$, Pawe\l{} Horodecki$^{(3)}$, Jonathan Oppenheim$^{(4)}$}





\begin{abstract}
We develop a formalism for distilling a classical key from a quantum state
in a systematic way, expanding on our previous work on secure key from bound entanglement [K. Horodecki {\it et. al.}, Phys. Rev. Lett. 94 (2005)].
More detailed proofs, discussion and examples are provided of the main results.  Namely, we demonstrate that
all quantum cryptographic protocols can be recast in a way which looks like entanglement theory, with
the only change being that instead of distilling EPR pairs, the parties distill private states. 
The form of these
general private states are given, and we show that there are a number of useful ways of expressing them. 
Some of the private states can be approximated by certain states which are bound entangled.  Thus distillable entanglement is not
a requirement for a
private key.  We find that such bound entangled states are useful for a cryptographic primitive we call a 
controlled private quantum channel.  We also find a general class of states which have negative partial
transpose (are NPT), but which appear to be bound entangled.
The relative entropy distance is shown to be an upper bound on the rate of key. This allows us to compute the {\it exact} value of distillable key for a certain class of private states. 
\end{abstract}


\tableofcontents
\section{ Introduction}

We often want to communicate with friends or strangers in private. 
Classically, this is impossible if we wish to communicate over long distances,
unless we have met before with our friend
and exchanged a secret key which is as long as the message we want to
send.  On the other hand, quantum
cryptography allows two people to communicate privately with only a very
short key which is just used to authenticate the message.

Every quantum cryptographic protocol is equivalent to the situation where
both parties (Alice and Bob) share some quantum state $\rab$, and then perform local operations on
that state and engage in public communication (LOPC) to obtain a key which is private from
any eavesdropper.
Until recently, every quantum protocol was also equivalent to distilling pure entanglement
from this shared state.  I.e., achieving privacy was equivalent to the two parties
converting many copies of the state $\rab$, to a smaller number of pure EPR pairs \cite{epr}
\be
|\psi_+ \>= {1\over \sqrt{2}}(|00\> + |11\>),
\label{eq:singlet}
\ee
using local operations and classical communication (LOCC), and then performing a measurement on
the EPR pairs in the computational basis.  Examples of such protocols include
BB84 \cite{BB84,ShorPreskill}, B92 \cite{B92,Tamaki03:secB92}, and of course, E91 \cite{Ekert91}.
It was thus thought that achieving security is equivalent to distilling pure entanglement,
and a number of results pointed in this direction 
\cite{QPA,DevetakWinter-hash,GisinWolf_linking,renner-wolf-gap,ALG_2_equiv,GisinWolf_QKAvsCKA,QKA_CKA_equivalance}.

Recently, however, we have shown that this is not the case -- there exist examples
of {\it bound entangled} states which can be used to obtain a secret key \cite{pptkey}.
Bound entangled states \cite{bound} are ones which need pure entanglement to create,
but no pure entanglement can be distilled from them. 
This helps explain the
properties of bound entangled states.  They have entanglement which protects correlations from
the environment (or an eavesdropper), but the entanglement  is so twisted that it can't be brought
into pure form.
This then raised the question of
what types of quantum states provide privacy.  In \cite{pptkey} we were able to find the 
general form of private quantum states $\gamma_{ABA'B'}$.  This allowed us to recast the theory of privacy
(under local operations and public communication -- or LOPC) in terms of entanglement 
theory (local operations and classical communication -- or LOCC).
In entanglement theory, the basic unit is the EPR pair, while in privacy theory, the only
difference is that one replaces the EPR pair with general {\it private states} $\gamma_{ABA'B'}$ as the basic units.

In the present article, we review the results of \cite{pptkey} in greater detail, and expand on the proofs and
tools.  Namely, we study and show that
the general form of a private state on a Hilbert space 
$\hcal_A\otimes\hcal_{A'}\otimes\hcal_B\otimes\hcal_{B'}$
with dimensions $d_A=d_B\equiv d$, $d_{A'}$ and $d_{B'}$,
is of the form 

\beq
\gamma_{ABA'B'} = 
U P^{+}_{AB}\otimes\sigma_{A'B'} U^\dagger, \nonumber
\eeq
where $P^{+}_{AB}$ is a projector onto the maximally entangled state 
$\psi_+=\sum_{i}{1\over \sqrt{d}} |e_i f_i\>$, 
and $U$ is the arbitrary {\it twisting operation} 
\be
U=\sum_{k,l=0}^{d-1} |e_k f_l\>\<e_k f_l|_{AB}\otimes U_{A'B'}^{kl} \s .
\nonumber
\ee
The key is obtained after measuring in the  $|e_i f_i\>$ basis.  We will henceforth refer to $\psi_+$
as the maximally entangled or EPR state (or Bell state in dimension $2\times 2$).
We show that the rate of key $K_D$ which can be obtained from a quantum state can be strictly greater
than the distillable entanglement, and this even holds if the distillable entanglement is strictly zero.
We also show \cite {pptkey} that
the size of the private key is generally bounded from above
by the regularized relative entropy of entanglement $E_r^\infty$ \cite{PlenioVedral1998}.  This will be
sufficient to prove that one can have a maximal rate of key strictly less than the entanglement cost 
(the number of singlets required to prepare a state under LOCC).

In section
\ref{sec:security} we introduce some of the basic concepts and terminology we will
use throughout the paper.  This includes the notion of {\it private states}, {\it pbits} which
contain one bit of private key, and {\it pdits} which have many bits of key.  In Section
\ref{sec:charact} we show that a state is secure if and only if it is of the form given above.  Then
we show different useful ways to write the private states in Section \ref{sec:pditprops},
and give some useful examples, and examine some of their properties.  This includes the notion
of {\it irreducibility} which is used to define the basic unit of privacy for private states.

States which have a perfect bit of key must have some distillable entanglement
\cite{AH-pditdist}.  The case of bound entangled
states with secure key is only found in the case of states 
which are not perfectly secure, although they are arbitrarily secure.  This motives our investigation
in Section \ref{sec:apbits} of {\it approximate pbits}.  We then demonstrate how to rewrite a bipartite state in terms of the eavesdropper's density matrix in Section \ref{sec:eve-states}.  This allows us
to interpret previous results in terms of the eavesdropper's states.  Then, in Section \ref{sec:overview},
we summarize the previous results in preparation for showing that bound entangled states can have a key.
In Sections \ref{sec:def-rate-key} and \ref{sec:equality}, we review the paradigms of entanglement (LOCC)
and privacy theory (LOPC), and show the equivalence of key rates in the two paradigms.  We then discuss
and compare security criteria in these paradigms in Section \ref{sec:securitycrit} of the Appendix.  In Section \ref{sec:bound} we
give a number of bound entangled states and show that they can produce a private key.  The methods allow
one to find a wide class of states which are bound entangled, because the fact that they have key automatically
ensures that they are entangled, which is usually the difficult part in showing that a state is bound entangled
(the PPT criteria can be quickly checked to see that the states are non-distillable).  In Section \ref{sec:relent}
we prove that the relative entropy distance is an upper bound on the rate of key.

In Section \ref{sec:nptbound}, a class of NPT states are introduced which appear to be bound entangled.
They are derived from a class of bound entangled private key states.
An additional result discussed in Section \ref{sec:cpqc}, which we only mentioned in passing in \cite{pptkey},
is that the bound entangled key states can be used
as the basis of a cryptographic primitive we call a controlled private quantum channel.
We conclude in Section \ref{sec:conc}
with a few open questions.

\section{Security contained in quantum states}
\label{sec:security}

In this section we will introduce the class of states which contain at least one bit (or dit) of perfectly secure key which is directly accessible -- these we call private bits (or dits). 
We discuss the properties of these states and argue the generality of this approach. In particular we introduce the notion of twisting, which is a basic concept in dealing with private states.

A well known state that contains one bit of secure key which is directly accessible is the singlet state. After measuring it in a local basis Alice and Bob obtain bits that are perfectly correlated with each other and completely uncorrelated with the rest of the world including an eavesdropper Eve. This is because the  singlet state as a whole is decoupled from the environment, being a pure state. However even if a state is mixed, it can contain secure key. Yet the key must then be located only in a part of it. More formally, we consider a four-partite mixed state  $\rh$ of two systems $A,A'$ belonging to Alice 
and $B,B'$ belonging to Bob. The $AB$ subsystem of the state will be called the {\it \main\ }
of the state -- it is the part of the state which produces key upon measurement.
The $A'B'$ subsystem will be called the {\it \side\ } of the state.  It is called this, because
its presence is what will cause the $AB$ part of the state to be secure, by {\it shielding} 
information from an eavesdropper. 

We assume the worst case scenario -- that the state is the reduced density matrix 
of the pure state $\psi_{ABA'B'E}$ where we trace out the system $E$ belonging to eavesdropper Eve. 
We then distinguish a product basis $\bcal=\{e_i,f_j\}$ in system $AB$. 
For our purposes, without loss of generality, we often choose $\bcal$ 
to be the standard basis $\{|ij\>\}$.   
Distinguishing the basis is connected with the fact that we 
are dealing with classical security, which finally is realized  in some fixed basis. 
Now, consider the state of systems $ABE$ {\it after } measurement 
performed in the basis $\bcal$ by Alice and Bob. This state is of the form 
\be
\rho_{ccq}=\sum_{i,j=0}^{d-1}p_{ij}|e_if_j\>_{AB}\<e_if_j|\otimes\rho_{ij}^E.
\label{eq:ccq-def}
\ee
The above form of the state is usually called a \ccq state. 
We will therefore refer to a \ccq state associated with state $\rh$,
and it is understood that it is also related to chosen basis $\bcal$. 
The distribution $p_{ij}$ will sometimes be referred to as the {\it distribution 
of the \ccq state}.

We can now distinguish types of states $\rh$ via looking at their \ccq states
(always assuming  that some fixed basis $\bcal$ was chosen):

{\defin  A state $\rho_{ABA'B'}$ is called {\bf secure} with respect to a
 basis  ${\cal B}\equiv{\{|e_i f_j\>_{AB}\}_{i,j=1}^d}$ if the state obtained via measurement 
 on AB subsystem of its purification in basis $\cal B$ followed by tracing out $A'B'$ subsystem (i.e. its \ccq state) is product with Eve's subsystem: 
\be
\bigl (\sum_{i,j=0}^{d-1} p_{ij}|e_i f_j\>\<e_i f_j|_{AB}\bigr )\otimes \rho_E.
\label{eq:ccq-secure}
\ee 
Such a state $\rho_{ABA'B'}$ will be also  called "${\cal B}$ secure".
Moreover if the distribution $\{p_{ij}\}=\{{1 \over d}\delta_{ij}\}$ 
so that the \ccq state is of the form
\be
\bigl (\sum_{i=0}^{d-1} {1\over d}|e_i f_i\>\<e_i f_i|_{AB}\bigr )\otimes \rho_E,
\label{eq:ccq-key}
\ee
the state $\rho_{ABA'B'}$ is said to have $\bcal$-key.
\label{bsec}
}

One can ask when two states $\rh$ and $\saabb$ are equally secure with respect 
to a given product basis $\bcal$.  First let us define what does it mean "equally secure".
A natural definition would be that when Alice and Bob measure systems AB
in the basis, then  Eve by any means cannot distinguish between  two situations, as far 
as the outcomes of the measurement are concerned. 
In particular, the states are definitely equally secure, 
when their \ccq states are equal.

For our purpose we will need to know when for two states 
the latter relation holds. It is obvious that any unitary transformation 
applied to systems $A'B'$ of the state $\rh$ will not change the ccq state. 
(Note that it cannot be just any CP map; for example, partial trace of 
systems $A'B'$ would mean giving it to Eve, which of course would change 
the ccq states). As it will be demonstrated in the next section, we can actually do much more without changing the ccq state. 
Namely we can apply an operation called "twisting". This operation 
is defined for system $ABA'B'$ and with respect to a product basis $\bcal$ of $AB$ 
system as follows:


{\defin
Given product basis $\bcal=\{e_i,f_j\}_{k,l}$ on systems $AB$, the unitary 
operation acting on system $ABA'B'$ of the form 
\be
U=\sum_{k,l=0}^{d-1} |e_k f_l\>\<e_k f_l|_{AB}\otimes U_{A'B'}^{kl},
\label{eq:twisting}
\ee
is called {\it $\bcal$-twisting}, or shortly {\it twisting}.
}

Finally we define the class of private states. The states from that class are proven \cite{pptkey}
to be the only quantum states which after measurement on Alice and Bob subsystems give
an ideal key. In other words these are the only states from which Alice and Bob can get an
ideal ccq state (\ref{eq:ccq-key}) according to definition \ref{bsec} of security. For the sake of
clarity, we recall this proof  with details in Section \ref{sec:charact}.

{\defin
\label{def:p_states} 
A state $\rh$ of a Hilbert space 
$\hcal_A\otimes\hcal_{A'}\otimes\hcal_B\otimes\hcal_{B'}$
with dimensions $d_A=d_B\equiv d$, $d_{A'}$ and $d_{B'}$,
of the form
\be
\gamma^{(d)} = {1\over d} \sum_{i,j=0}^{d-1} |e_i f_i\>\<e_j f_j|_{AB}\otimes U_i\sigma_{A'B'} 
U^{\dagger}_j,
\label{eq:ipdit}
\ee
where the state $\sigma_{A'B'}$ is an arbitrary state of subsystem $A'B'$, 
$U_i$'s are  arbitrary unitary  transformations 
and $\{|e_i\>\}_{i=0}^{d-1}$,\,$\{|f_j\>\}_{i=0}^{d-1}$ are local basis on $\hcal_A$ and $\hcal_B$ respectively, is called {\bf private state} or {\bf \pdit}.
In case of $d=2$ the state is called {\bf \pbit}.
}

Note, that maximally entangled states are also private states, which is in case when $d_A' = d_B' =1$. In general, any \pdit\ can be created out of a maximally entangled  state with additional state on $\sigma_{A'B'}$ 
(which we will call {\it basic \pdit}) by some twisting. 

{\defin  A state $\rh$ of a Hilbert space 
$\hcal_A\otimes\hcal_{A'}\otimes\hcal_B\otimes\hcal_{B'}$
with dimensions $d_A=d_B\equiv d$, $d_{A'}$ and $d_{B'}$, of the form
 
\be 
\rh = P^{+}_{AB}\otimes\sigma_{A'B'}, 
\label{eq:ibasicpdit}
\ee
is called a {\bf basic \pdit}.
}

{\rem Let us note, that one could define states with ideal key also in a different way than
in definition \ref{bsec}.
Namely, we could say that the state $\rho_{AB}$ has $\bcal$ key iff it has a subsystem $ab$ 
such, that the subsystem $abE$ of its purification $|\psi_{ABE}\>$ is an ideal ccq state of eq. (\ref{eq:ccq-key}), 
where $\bcal = \{|e_i f_i\>\}_{i=1}^{d}$. However maximally entangled state is not of this form,
but is locally equivlent to a state of this form, i.e. they would be transformable into one another 
by means of unitary embeddings or partial isometries. In fact the whole class of so defined states with ideally 
secure key, would be locally equivalent to the one we have introduced in definition \ref{def:p_states}, and by
characterization of the latter, equivalent to the class of private states. Another definition of states with 
ideal key could be as follows. A state $\rho_{AB}$ is called to have key if there are operations $\Lambda_A$ 
and $\Lambda_B$ with Kraus operators $\{|i\>_{K_A}\ot X^{(i)}_A\}$ and $\{|i\>_{K_B}\ot Y^{(i)}_B\}$ respectively, 
such that the $ccq$ state of subsystems $K_AK_BE$ of the purification of an output 
state $\Lambda_A\ot\Lambda_B(\rho_{AB})$, is an ideal ccq state of eq. (\ref{eq:ccq-key}). 
 This definition however would not allow for easy characterization of this class of states, 
and still, such states would be locally equivalent to private states defined in \ref{def:p_states}.
}
\subsection{Some facts and notations}
\label{subsec:notation}
In what follows, by $||.||$ we mean the trace norm, i.e. the sum of the singular values of an operator. For any bipartite operator $X \in B(\hcal_{1}\ot\hcal_{2})$, by $X^{\Gamma}$ we mean the partial transposition of $X$ with respect to system 2, that is:
\be
(\id_1\ot T_2)X,
\ee
where $T_2$ denotes the matrix transposition over system $2$ of a matrix $X$. For brevity, we will use the same symbol for any partial transposition. In particular, we deal often with systems of four subsystems $ABA'B'$, and we will take partial transposition with respect to subsystems $B$ and $B'$. Hence $\rho_{ABA'B'}^{\Gamma}$ denotes $(\id_A\ot T_B\ot\id_{A'}\ot T_{B'})(\rho_{ABA'B'})$, with $T_B$ and $T_{B'}$ denoting the matrix transposition on systems $B$ and $B'$ respectively. To give example of a mixed notation, we consider $\rho_{ABA'B'} \in B(\ccal^2\ot\ccal^2\ot\ccal^d\ot\ccal^d)$. In block matrix form, such a state has a bipartite structure of blocks, so that it reads:
\be
\rho_{ABA'B'}=\left[\bea{cccc}
A_{0000} &A_{0001}&A_{0010}&A_{0011} \\
A_{0100}& A_{0101}&A_{0110}&A_{0111} \\
A_{1000}&A_{1001}&A_{1010}& A_{1011}\\
A_{1100} &A_{1101}&A_{1110}& A_{1111}  \\
 \eea
\right].
\label{blockform_0}
\ee
This state after partial transposition with respect to system $BB'$ reads:
\be
\rho_{ABA'B'}^{\Gamma}=\left[\bea{cccc}
A_{0000}^{\Gamma} &A_{0100}^{\Gamma}&A_{0010}^{\Gamma}&A_{0110}^{\Gamma} \\
A_{0001}^{\Gamma}& A_{0101}^{\Gamma}&A_{0011}^{\Gamma}&A_{0111}^{\Gamma} \\
A_{1000}^{\Gamma}&A_{1100}^{\Gamma}&A_{1010}^{\Gamma}& A_{1110}^{\Gamma}\\
 A_{1001}^{\Gamma}&A_{1101}^{\Gamma}&A_{1011}^{\Gamma}& A_{1111}^{\Gamma}  \\
 \eea
\right].
\ee
In the above equation, the partial transposition on LHS is with respect to system $BB'$ and 
on the RHS, only with respect to system $B'$, as the partial transposition with respect to
system $B$, which is a one qubit system, resulted already in appropriate reordering of the block operators $A_{ijkl}$.

In what follows, we will repeatedly use equivalence of the trace norm distance and fidelity proved by Fuchs  and van de Graaf \cite{Fuchs-Graaf}:
{\lemma 
For any states $\rho$, $\rho'$ there holds
\be
1-F(\rho,\rho')\leq {1\over 2}||\rho - \rho'||\leq \sqrt{1-F(\rho,\rho')^2}
\label{fidnor}
\ee
Here $F(\rho,\rho')=\tr \sqrt{\sqrt\rho \rho' \sqrt\rho}$ is fidelity; 
\label{lemfidnor}
}

We also use the Fannes inequality \cite{Fannes1973} (in the form of \cite{Alicki-Fannes}):
\be
|S(\rho) - S(\sigma)|\leq 2||\rho-\sigma||\log d + h(||\rho - \sigma||),
\label{eq:Fannes_ineq}
\ee
which holds for arbitrary states $\rho$ and $\sigma$
satisfying $||\rho - \sigma || \leq 1$.

\subsection{On twisting and \cohsq}
\label{ssec:twistcosq}
Here we will show that twisting does not change the \ccq state
arising from measurement of the \main. Then we will 
introduce a useful tool by showing that twisting 
can pump entanglement responsible for security of \ccq state
into the \main. 

We have the following theorem. 

  \vskip0.3cm
{\theorem 
\label{th:ccq}
For any state $\rho_{AA'BB'}$ and any $\bcal$-twisting operation $U$, the states 
$\rho_{AA'BB'}$ and $\saabb=U\rho_{AA'BB'}U^{\dagger}$ 
have the same ccq states w.r.t $\bcal$, i.e. after measurement 
in basis $\bcal$, the corresponding ccq states are equal: 
$\tilde \rho_{ABE}=\tilde \sigma_{ABE}$}

{\proof  To show that subsystem $\rho_{ABE}$ is not affected by $\bcal$ controlled unitary 
with a target on $A'B'$ we will consider the whole pure state: 
\be
|\psi_{ABA'B'E}\>= \sum_{ijklm} a_{ijklm}|ijklm\>\equiv |\psi\>
\ee
(without loss of generality we take $\bcal$ to be standard basis). 
After von Neumann measurement on $\bcal$ and tracing out the $A'B'$ 
part the output state is the following:
\be
\tilde{\rho}_{ABE}= \sum_{ijklmn}a_{ijklm}\bar{a}_{ijkln} |ij\>\<ij|\otimes|m\>\<n|.
\label{parttrace}
\ee 
Let us now subject $|\psi\>$ to controlled unitary $U_{ABA'B'}\otimes I_E$,
\be
U_{ABA'B'}\otimes I_E|\psi\>=\sum_{ijklm} a_{ijklm}|ij\>U^{ij}|kl\>|m\>\equiv |\tilde{\psi}\>,
\ee
and then on the output state $|\tilde{\psi}\>$ perform a complete measurement 
on $\bcal$ reading the output:
\ben
 P_{ij} |\tilde{\psi}\>\<\tilde{\psi}| P_{ij} = \sum_{klmstn} a_{ijklm} \bar{a}_{ijstn} 
\nonumber \\  
|ij\>\<ij|_{AB}\otimes U^{ij}|kl\>\<st|(U^{ij})^{\dagger}_{A'B'}\otimes |m\>\<n|_E.
\een
Performing partial trace and summing over $i,j$ we obtain 
the same density matrix as in (\ref{parttrace}) 
which ends the proof. 
} \blacksquare

The above theorem states that two states which differ by some twisting $U$, have the 
same $ccq$ state obtained by measuring their \mains, and tracing out their \sides.
However, since twisting does not affect only the \ccq state, one can 
be interested in how the whole state changes when subjected to such an operation. 
We will show now an example of twisting which will be of great importance 
for further considerations in this paper. Subsequently, we will construct from this 
twisting an operation called {\it \cohsq}
(shortly: {\it \cosq }), which shows the importance of the above theorem. 
The operation of \cosq\ is a kind of primitive in the paradigm which we 
will present in the paper. 

Consider the following technical lemma:
{\lemma 
For any state $\sigma_{ABA'B'} \in {\cal B}({\cal C}^2\ot{\cal C}^2 \ot {\cal C}^d\ot{\cal C}^{d'})$ 
expressed in the 
form $\sigma_{ABA'B'} = \sum_{ijkl=0}^{1} |ij\>\<kl|\ot A_{ijkl}$ there exists twisting 
$U_{ps}$ such that if we apply this to $\sigma_{ABA'B'}$, and trace out $A'B'$ part, 
the resulting state on $AB$  $\rho_{AB}=\tr_{A'B'}[U_{ps}\sigma_{ABA'B'}U_{ps}^{\dagger}]$ will 
have the form
\be
\rho_{AB}=
\left[\bea{cccc}
\times & \times &\times & ||A_{0011}||\\
\times & \times &\times &\times \\
\times & \times & \times  &\times \\
\times & \times & \times &\times \\
\eea\right],
\label{eq: rho:upbig}
\ee
where we omit  non-important elements of $\rho_{AB}$.
\label{lem:secpar}
}

{\proof Twisting, by its definition (\ref{eq:twisting}) is determined by the set of 
unitary transformations. In the case of \pbit\ which we now consider, there are
four unitary transformations which  determine it: 
$\{ U_{kl}\}_{k,l=0}^{1}$.  Let us consider singular value
decomposition of the operator $A_{0011}$ to be $VR \tilde V$ with $V,\tilde V$ unitary 
transformations, and $R$ - nonnegative diagonal operator. Note, that by 
unitary invariance of norm, we have that 
$||A_{0011}|| = ||R|| = \tr R$. We then define twisting by choosing $U_{00}=V^{\dagger}$, 
$U_{11}=\tilde V$, and $U_{01}=U_{10}=I$.  
The $AB$ subsystem of twisted $\sigma_{ABA'B'}$ state is
\be
\rho_{AB}=\sum_{ijkl=0}^1  \tr (U_{ij}A_{ijkl}U_{kl}^{\dagger})|ij\>\<kl|,
\ee   
so for such chosen twisting we have indeed, that the element $|00\>\<11|$ of the matrix of $\rho_{AB}$  is
equal to $\tr  U_{00}^{\dagger}V R {\tilde{V}} U_{11}^{\dagger}= \tr R = ||A_{0011}||$, which 
proves the  assertion.
}\blacksquare

We will give now the following corollary, which will serve as simple 
exemplification of this result.

{\corrolary  Let the \main\ be two qubit system. 
Consider then a state of the form (where blocks are operator 
acting on $A'B'$ system):
\be
\sigma_{ABA'B'}=\left[\bea{cccc}
A_{0000} &0&0&A_{0011} \\
0& A_{0101}&A_{0110}&0 \\
0&A_{1001}&A_{1010}& 0\\
A_{1100} &0&0& A_{1111}  \\
 \eea
\right],
\label{blockform}
\ee
there exists twisting such that
the state after partial trace on $A'B'$ has a form
 \be
\rho_{AB} = \left[\bea{cccc}
||A_{0000}|| &0&0&||A_{0011}|| \\
0& ||A_{0101}||&||A_{0110}||&0 \\
0&||A_{1001}||&||A_{1010}||& 0\\
||A_{1100}|| &0&0& ||A_{1111}||  \\
\eea
\right].
\label{eq:normform}
\ee
}

{\proof The construction of the twisting is similar as in lemma above. This time one 
has to consider also
the singular value decomposition of the operator  
$A_{0110} = W S W'$. } \blacksquare

We can see now, that with any state $\rho_{ABA'B'}$, which has 
two qubit  \main\  $AB$,  we can associate 
a state obtained in the following way:
\begin{enumerate}
\item For state $\rho_{ABA'B'}$ find twisting $U_{ps}$, such, 
that (according to lemma \ref{lem:secpar}) it changes upper-right element of $AB$
subsystem of $\rho_{ABA'B'}$ into $||A_{0011}||$.
\item Apply $U_{ps}$ to $\rho_{ABA'B'}$ obtaining $\rho'_{ABA'B'}=U_{ps} \rho_{ABA'B'}U_{ps}^{\dagger}$.
\item Trace out the \side\ ($A'B'$ subsystem) of state $\rho'_{ABA'B'}$ obtaining two-qubit state 
\be
\rho'_{AB}=\tr_{A'B'}\rho'_{ABA'B'}.
\ee
\end{enumerate}
This operation we will call {\bf \cohsq\ }, or shortly {\bf \cosq\ }, 
and the state $\rho'_{AB}$ which is the output of such operation on the 
state $\rho_{ABA'B'}\in \bcal (\ccal^2\ot\ccal^2\ot\ccal^d\ot\ccal^{d'})$ the 
{\bf \cosqd\ state of } the state $\rho_{ABA'B'}$. Sometimes we shall use the term \cohsq\ in
more informal sense, namely, with the twisting $U_{ps}$ which makes the key 
part close to maximally entangled state.

Note, that the \ccq state of \cosqd\ state 
has no more secret correlations than that  of the original state.
This is because it emerges from the operation of twisting which 
preserves security in some sense, i.e. it does not change
the \ccq state which can be obtained from the original state. The next operation 
performed in definition of \cosqd\ 
state is tracing out $A'B'$ part which means giving the $A'B'$ subsystem 
to Eve. Such operation can not increase
security of the state in any possible sense.


We will be interested  in applying \cosq\ in the 
case, where the \main\ of the initial state 
was weakly entangled, or completely separable. 
Then the \cosq\ operation will make it entangled.

We can say, that the operation of \cohsq\ pumps the entanglement of 
the state which is distributed along subsystems $AA'BB'$ into its \main\ $AB$. 
The entanglement once concentrated  
in the two qubit part, may be much more powerful than
the one spread over the whole system. Further in the paper,
we will see that from the bound entangled state,
the operation of \cosq\ can produce approximately a maximally entangled state 
of two qubits.  Then the analysis of how much key one can draw from 
the \ccq state is much easier in the case of the \cosq\ state.



\section{General form of states containing ideal key.}
\label{sec:charact}
In this section we will provide general form of the states $\rh$ 
which have $\bcal$ key, i.e. states such that the outcomes of measurement 
in basis $\bcal$ are both perfectly correlated and perfectly secure.
It turns out that this is precisely the class of private states. Hence
the definitions \ref{bsec} and \ref{def:p_states} are equivalent. We have the following theorem:

{\theorem Any state $\rh$ of a Hilbert space 
$\hcal_A\otimes\hcal_{A'}\otimes\hcal_B\otimes\hcal_{B'}$
with dimensions $d_A=d_B\equiv d$, $d_{A'}$ and $d_{B'}$,
has $\bcal$-key if and only if it is of the form 
\be
\rh = {1\over d} \sum_{i,j=0}^{d-1} |e_i f_i\>\<e_j f_j|_{AB}\otimes U_i\sigma_{A'B'} 
U^{\dagger}_j
\label{eq:pdit}
\ee
where the state $\sigma_{A'B'}$ is an arbitrary state of subsystem $A'B'$, 
$U_i$'s are  arbitrary unitary  transformations 
and $\{e_i\otimes f_j\}=\bcal$.
\label{th:pdit}
}

We can rewrite the state (\ref{eq:pdit}) in the following, more appealing form 
\be
\rh=U P^{+}_{AB}\otimes\sigma_{A'B'} U^\dagger,
\label{eq:pdit2}
\ee
where $P^{+}_{AB}$ is a projector onto the maximally entangled state 
$\psi_+=\sum_{i}{1\over \sqrt{d}} |e_i f_i\>$, 
and $U$ is arbitrary twisting operation (\ref{eq:twisting}). 
Since the state $P_{AB}^+$ has many matrix elements vanishing, not all unitaries 
from definition of twisting are actually used here. In fact, 
unitaries $U_i$ in equation (\ref{eq:pdit}) are to be identified 
with unitaries $U_{kk}$ from equation (\ref{eq:twisting}). 
Note, that we can take $\sigma_{A'B'}$ to be  "classically correlated" in the 
sense that it is diagonal in some product basis. Indeed, twisting 
can change the state $\sigma_{A'B'}$ into any other state having the same eigenvalues
(simply, twisting can incorporate a unitary transformation acting solely on $A'B'$).

Thus we see that the states  which  have key, are closely connected with the maximally entangled state,
which has been so far a "symbol" of quantum security. As we shall see,
the maximally entangled state may get twisted so much,  that after measurement in many 
bases of the $AB$ part  the outcomes will be  correlated
with Eve, which is not the case for the maximally entangled state itself.  
Still, however the basis $\bcal$ will remain secure.
Note that here we deal with perfect security.  We will later discuss approximate
security in Section \ref{sec:apbits}.


{\proof  ($\Leftarrow$)

This part of the proof is a consequence of the theorem \ref{th:ccq}. 
Namely a {\it basic \pdit\ } (\ref{eq:ibasicpdit}) is obviously $\bcal$-secure, 
because it has maximal correlations  in this basis, and 
moreover it is a pure state, hence the one completely decoupled from Eve.
More formally, it is evident that the  \ccq state of basic \pdit\  is of the form 
(\ref{eq:ccq-key}).
Now we can apply  theorem \ref{th:ccq},
which says that after twisting the \ccq state is unchanged. Hence 
any state of the form (\ref{eq:pdit}) has also $\bcal$ key.  
}

{\proof ($\Rightarrow$)

In this part we assume, that the state $\rh$ has $\bcal$-key i.e. 
that after measurement on it's $AB$ part,
one gets a perfectly correlated state (between Alice and Bob) that is uncorrelated with Eve:
\be
\bigl (\sum_{i=0}^{d-1} {1\over d}|e_i f_i\>\<e_i f_i|_{AB} \bigr )\otimes \rho_E.
\label{eq:wanted}
\ee
Let us consider general pure state for which dimensions of $A, B$ are $d$, 
dimensions of $A', B'$ are $d_{A'},d_{B'}$ respectively, and dimension of subsystem $E$ is the smallest one which allows for the whole state being a pure one. 
\be
 |\psi\> =|\psi_{ABA'B'E}\>= \sum_{ijklm} a_{ijklm}|e_i f_j klm\>.
\ee
one can rewrite it as
\be
|\psi\> = \sum_{ij}|e_i f_j\>_{AB}|\tilde \psi^{(ij)}\>_{A'B'E}.
\label{vec}
\ee
with $|\tilde \psi^{(ij)}\>_{A'B'E} = \sum_{klm} a_{ijklm} |klm\>$.

It is easy to see that 
the scalar product $\<\tilde \psi^{(ij)}|\tilde \psi^{(ij)}\>$ equals the probability
of obtaining the state $|e_i f_j\>\<e_i f_j|_{AB}$ on the system $AB$ after measurement in
basis $\bcal$. Now, since the subsystem $\rho_{ABE}$ (after measurement in $\bcal$ on AB)
must be maximally correlated, the vectors $|\tilde \psi^{(ij)}\>$ 
should satisfy $\<\tilde \psi^{(ij)}|\tilde \psi^{(ij)}\>={1\over d}\delta_{ij}$. 
We can normalize these states (in case $i =j$) to have:
\be
|\psi^{(ii)}\>:= {|\tilde\psi^{(ii)}\>\over {\sqrt{\<\tilde \psi^{(ii)}|\tilde \psi^{(ii)}\>}}} =\sqrt{d}|\tilde{\psi}^{(ii)}\>
\ee
so that the total state has a form:
\be
|\psi\>= \sum_{i=0}^{d-1} {1\over \sqrt{d}}|e_i f_i\>_{AB}|\psi^{(ii)}\>_{A'B'E}.
\ee
"Cryptographical" interpretation of this state is the following: if Alice and Bob gets 
$i-$th result, then Eve gets subsystem $\rho_i^E$ of a state 
$|\psi^{(ii)}\>_{A'B'E}$. 
Indeed, the ccq state is then of the form
\be
\rho_{ccq}=\sum_{i=0}^{d-1}{1\over d}|e_if_i\>_{AB}\<e_if_i|\otimes\rho_{i}^E,
\ee
with $\rho_{i}=\tr_{A'B'}(|\psi^{(ii)}\>_{A'B'E}\<\psi^{(ii)}|)$.
Now the condition (\ref{eq:wanted}) implies that, 
$\rho_i^E$ should be all equal to each other.
In particular, it follows that rank of Eve's total density 
matrix is no  greater than dimension of $A'B'$ system, 
hence we can assume that $d_E=d_{A'}d_{B'}=d'$. 
It is convenient to rewrite this pure state in a form 
\be
|\psi^{(ii)}\>_{A'B'E}=\sum_{k=0}^{d'-1} |k\>_{A'B'}X_i|k\>_{E},
\label{eq:psi-ii}
\ee
where $\{|k\>\}$ is  standard basis of $A'B'$ and of $E$ system,
$X_i$ is  $d_E\times d_E$ matrix that fully represents this state. 
It is easy to check,  that $\rho^E_i=X_iX_i^{\dagger}$.  Consider now  singular
value decomposition of $X_i$ given by $V_i \sqrt{\rho_i} U_i^{\dagger}$
where $\rho_i$ is now diagonal in basis $\{|k\>\}$. 
One then gets that  $\rho^E_i=V_i \rho_i V_i^{\dagger}$. 
The state (\ref{eq:psi-ii}) may be rewritten
\be
|\psi^{(ii)}\>_{A'B'E}=\sum_k X_i^T|k\>_{A'B'}|k\>_{E},
\ee
where $T$ is transposition in basis  $\{|k\>\}$.
Now it is easy to check, that subsystem 
$A'B'$ of $|\psi^{(ii)}\>_{A'B'E}$ is in state $X_i^T (X_i^T)^{\dagger}$, so that
the whole state $\rho_{ABA'B'}$ is the following:
\be
\rho_{ABA'B'} = 
{1\over d}\sum_{i,j=0}^{d-1}|e_i f_i\>\<e_j f_j|_{AB}\otimes X_i^T (X_j^{\dagger})^T.
\label{eq:pdit-X}
\ee
We can express this state using states accessible to Eve, namely $\rho^E_j$:
\ben
\rho_{ABA'B'} = 
{1\over d}\sum_{i,j=0}^{d-1}|e_i f_i\>\<e_j f_j|_{AB}\otimes  \nonumber \\
(U_i^{*}V_i^T) \underbrace{V_i^*\sqrt{\rho_i}^{T}V_i^T}_{={\sqrt{\rho_i^E}}^T} 
\underbrace{V_j^*\sqrt{\rho_j}^{T}V_j^T}_{={\sqrt{\rho_j^E}}^T} (V_j^*U_j^T).
\een
(For expressing state in terms of Eve's states in more general case, see 
Section \ref{sec:eve-states}).
Denoting by $W_i$ the unitary transformation $U_i^{*}V_i^T$ one gets:
\be
\rho_{ABA'B'} = 
{1\over d}\sum_{i,j=0}^{d-1}|e_i f_i\>\<e_j f_j|_{AB}\otimes  \nonumber \\
W_i {\sqrt{\rho_i^E}}^T . {\sqrt{\rho_j^E}}^T W_j^{\dagger}.
\ee
However, as mentioned above, Eve's density matrices are equal to each other, i.e. 
$\rho_i^E=\rho_j^E$ for all $i,j$. We then obtain 
\be
\rho_{ABA'B'} = {1\over d}\sum_{i,j=0}^{d-1}|e_i f_i\>\<e_j f_j|_{AB}\otimes 
W_i\rho {W_j^{\dagger}}_{A'B'}.
\ee
This completes the proof of theorem \ref{th:pdit}.
}\blacksquare



\section{Pdits  and their properties}
\label{sec:pditprops}

In this section we will present various forms of \pdits\ and \pbits.
We will first write the {\pbit\ }in matrix form according to its original definition. 
We can write it in block form 
\be
\gamt ={1\over 2}\left[\bea{cccc}
U_0 \sigma_{A'B'}U_0^{\dagger} &0&0&U_0 \sigma_{A'B'}U_1^{\dagger} \\
0& 0&0&0 \\
0&0&0& 0\\
U_1 \sigma_{A'B'}U_0^{\dagger}&0&0&U_1 \sigma_{A'B'}U_1^{\dagger}\\
\eea
\right],
\label{eq:standform}
\ee
where $\sigma_{A'B'}$ is arbitrary state on $A'B'$ subsystem, 
and $U_0$ and $U_1$ are arbitrary unitary transformations which act on $A'B'$.

{\it "Generalized EPR form" of pdit.} Since by the theorem of the previous section 
\pdits\  are 
the only states that contain $\bcal$-key,
they could be called generalized EPR states (maximally entangled state). We have already seen 
that they are "twisted EPR states". 
 One can notice an even closer connection. 
Namely, a \pdit\  can be viewed as an {\it EPR states with operator amplitudes}. Indeed,
one can rewrite equation (\ref{eq:pdit-X}) in a more appealing form
\be
\gamd = \Psi \Psi^\dagger,
\ee
with 
\be
\Psi={1\over \sqrt d}\sum_{i=0}^{d-1}Y_i^{A'B'}\otimes |e_i f_i\>_{AB}.
\ee
We have written here (unlike in the rest the of paper) first 
the ${A'B'}$ system and then the $AB$ one, so that this form of \pdit\
would recall a form of pure state. 
Thus instead of $c$-numbers the amplitudes are now $q$-numbers,
so that states which have key are "second quantized EPR states".
In the case of \pbits, the matrix form 
is  the following:
\be
\gamt={1\over 2}\left[\bea{cccc}
Y_0Y_0^{\dagger} &0&0&Y_0Y_1^{\dagger} \\
0& 0&0&0 \\
0&0&0& 0\\
Y_1Y_0^{\dagger}&0&0&Y_1Y_1^{\dagger}\\
\eea
\right].
\ee
Let us consider the polar decomposition of operators $Y_i$. From
definition of \pdit\ it follows that the only constraint on these
operators is the following 
\be
\forall_i \,\, Y_i = U_i\sqrt{\rho},
\ee
where $U_i$ is unitary transformation and $\rho$ is a normalized state
as so is the $\sigma_{A'B'}$ state in form (\ref{eq:standform}).


This reflects the fact, that
similarly like maximally entangled state which produces correlated 
outputs has amplitudes with
probably different phases, but the same moduli, the "maximally
private" state can have "operator amplitudes" 
which differ by $U_i$ ( a counterpart
of the phase) but have the same $\sqrt{\rho}$ in polar 
decomposition
(which is a counterpart of the modulus).

There is yet another
similarity to EPR states, namely the norm of 
upper-right block $Y_0Y_1^{\dagger}$ is equal to
${1\over 2}$, like the modulus of the coherence of the EPR state.

{\it  "$X$-form" of  \pbit.} In special case of  \pbits\  one can have representation by just 
one normalized operator:
\be
\gamt ={1\over 2}\left[\bea{cccc}
\sqrt{XX^{\dagger}} &0&0&X \\
0& 0&0&0 \\
0&0&0& 0\\
X^{\dagger}&0&0&\sqrt{X^{\dagger}X}\\
\eea
\right],
\label{eq:Xform}
\ee
for any operator $X$ satisfying $||X|| =1$.

Justification of equivalence of this form and standard form is the following.
Consider singular value decomposition of $X$
$X= U \sigma W$ with $U$ and $W$ unitary transformations and 
$\sigma$ being diagonal, positive matrix.
Since $X$ has trace norm 1, the same is for  $\sigma$, 
therefore it can be viewed as $X=U \rho W$ with $\rho$ being a legitimate state.
Identifying $U_0=U$ and $U_1=W^\dagger$ we obtain standard form.

It is important, that in nontrivial cases $X$ should be a  non-positive operator. 
Otherwise the \pbit\ is equal to basic \pbit. Indeed, if it is positive,
then since its trace norm is 1, it is itself a legitimate state, call it $\rho$. 
Then $\sqrt{XX^{\dagger}}=\sqrt{X^{\dagger}X} = \rho$,
so that 
$$\rho_{ABA'B'} = {1\over 2}\sum_{i,j=0}^1|ii\>\<jj|\ot 
\rho = |\psi_+\>\<\psi_+|\ot \rho.$$
which is the basic \pbit\ (\ref{eq:ibasicpdit}).

Note, that in higher dimension to have the $X$-form we need more than one operator,  
and the operators depend 
on each other, which is not as simple representation as in the case of \pbit. 
For example in $d=3$ case we have:
\ben
{1\over 3}\left[\bea{ccccccccc}
\sqrt{XX^{\dagger}} &0&0& 0 &X&0& 0&0&XY \\
0& 0&0&0& 0&0&0& 0&0 \\
0& 0&0&0& 0&0&0& 0&0\\
0& 0&0&0& 0&0&0& 0&0\\
X^{\dagger} &0&0& 0 &\sqrt{X^{\dagger}X}&0& 0&0&Y\\
0& 0&0&0& 0&0&0& 0&0\\
0& 0&0&0& 0&0&0& 0&0\\
0& 0&0&0& 0&0&0& 0&0\\
(XY)^{\dagger} &0&0& 0 &Y^{\dagger}&0& 0&0&\sqrt{Y^{\dagger}Y}\\
\eea
\right],
\een
where the operators $X$ and $Y$ satisfy: $||X||=1$ and $X = W Y^{\dagger}$ for arbitrary unitary transformation $W$.

{\it "Flags form": special case of  $X$-form.} If the operator $X$ which 
represents \pbit\ in its $X$-form is additionally hermitian, any 
such \pbit\ can be seen as a mixture of  {\it basic} \pbit\ 
and a variation of {\it basic} \pbit\ which has EPR states  with different phase:
\be
\gamt = p|\psi_+\>\<\psi_+| \ot \rho^+_{A'B'}+(1-p)|\psi_-\>\<\psi_-|\ot \rho^-_{A'B'},
\label{eq:flagform}
\ee
where $|\psi_\pm\>={1\over \sqrt{ 2}}(|00\> \pm |11\>)$.
Derivation of this form is straightforward, if we consider decomposition of X 
into positive and negative part \cite{Bhatia}:
\be
X= X_+ - X_-,
\ee
where $X_+$ and $X_-$ are by definition orthogonal, and positive. Thus 
denoting $p = \tr X_+$, together with assumption of $X$-form that 
$||X|| = \tr|X| = 1$, we can rewrite $X$ as
\be
X = p\rho_+ - (1-p)\rho_-,
\ee
where $\rho_\pm $ are normalized positive and negative parts of $X$. Moreover, since the 
states $\rho_+$ and $\rho_-$ are
orthogonal: $\tr\rho_-\rho_+ =0$, we obtain the form (\ref{eq:flagform}).

\subsection{Private bits - examples}
\label{ssec: pbtexpls}
We will give now two examples of private bits, and study its entanglement distillation properties.

{\it Examples of \pbit}
\begin{enumerate}
\item Let us consider state $\gamma^{V} 
\in B(\ccal ^2 \ot \ccal^2 \ot \ccal^d \ot \ccal^d)$ of the following form:
\be
\gamma^{V}={1\over 2}\left[\bea{cccc}
{\id \over d^2} &0&0&{V\over d^2} \\  
0& 0&0&0 \\
0&0&0& 0\\
{V\over d^2}&0&0&{\id \over d^2}\\
\eea
\right],
\label{eq:explpbit}
\ee
where $V$ is the swap operator which reads: $V=\sum_{i=0}^{d-1}|ij\>\<ji|$.
If we consider positive and negative part of $V$, which are symmetric and antisymmetric subspace, 
it is easy to see, that 
\be
\gamma^{V}= p|\psi_+\>\<\psi_+|\ot \rho_s  + (1-p)|\psi_-\>\<\psi_-|\ot \rho_a
\ee
where 
\be
\rho_s = {2\over d^2 +d}P_{sym}\quad \rho_a ={2\over d^2 -d}P_{asym}
\label{eq:Werner-sym-asym}
\ee
are symmetric and antisymmetric Werner states, and $p={1\over 2}(1+{1\over d}$).
Thus we have obtained, that it is also a \pbit\ with natural "flags form", with flags being (orthogonal) Werner 
states \cite{Werner1989}.
\item The second example is the state known as "flower state", which was 
shown \cite{lock-ent} to lock entanglement cost. We have that 
$\rho_{flower}  \in B(\ccal ^2 \ot \ccal^2 \ot \ccal^{d^2} \ot \ccal^{d^2})$ is of the form:
\be
\gamma_{flower}={1\over 2}\left[\bea{cccc}
\sigma &0&0&{1\over d}U^T \\  
0& 0&0&0 \\
0&0&0& 0\\
{1\over d}U^*&0&0&\sigma\\
\eea
\right],
\label{eq:explflow}
\ee
where $\sigma$ is classical maximally correlated state: $\sigma = \sum_{i=0}^{d-1}{1\over d}|ii\>\<ii|$, and $U$ 
is the embedding of unitary transformation $W= \sum_{i,j=0}^{d-1} w_{ij} |i\>\<j| = H^{\ot {\log d}}$ with $H$ being
Hadamard transform in the following way: $$ U = \sum_{i,j=0}^{d-1}w_{ij} |ii\>\<jj|.$$
We can check now, that this state is pbit with $X$-form. In this case $X = U^T$. To see this consider unitary transformation 
$S:=U^* + \sum_{i\neq j}|ij\>\<ij|$. Composing $S$ with $U^T$ does not change the norm, which is unitarily invariant,
so that 
\be
||{1\over d}U^T|| =||{1\over d}U^T S ||=||{1\over d}\sum_{i=0}^{d-1}|ii\>\<ii||| =1.
\ee
Thus we see, that $||X||=1$. We have also $\sqrt{XX^{\dagger}}=\sigma$:
\be
\sqrt{{1\over d^2}U^T U^*} = [{{1\over d^2} \sum_{i=0}^{d-1}|ii\>\<ii|}]^{1\over 2}=\sigma.
\ee
\end{enumerate}

We will show now, that in case of $\gamma^V$ given in eq. (\ref{eq:explpbit}),  
the distillable entanglement $E_D$  is strictly smaller then the amount of secure key $K_D$ gained 
from these states. The formal definition of $K_D$ is given in 
Section \ref{sec:def-rate-key}. Here it is enough to base only
on its intuitive properties. Namely, any \pdit\  by its very definition 
has $K_D$ at least equal to $\log d$ of key, 
which can be obtained by measuring its \main. To show the gap between 
distillable  entanglement and distillable key we will compute the value of another 
measure of entanglement:  log-negativity $E_N(\rho)$ (see \cite{ZyczkowskiHSP-vol})
of the state, which is an upper bound on distillable entanglement 
\cite{Vidal-Werner}. To this end consider the following lemma.
{\lemma For any \pbit\ in $X$-form, if $\sqrt{XX^\dagger}$ and $\sqrt{X^{\dagger}X}$ are PPT, 
the log negativity of the \pbit\ in $X$-form reads $E_N=\log (1 + ||X^{\Gamma}||)$, 
where ${\Gamma}$  is transposition performed on the system $B'$.
\label{lem:negofpbit}
}

The proof of this lemma is given in Sec. \ref{subsec:app_prop_of_pbits} of Appendix. Using this lemma, one can check the 
negativity of  the state $\gamma^V$. 
We have in this case $X = {V\over d^2}$, with $d\geq 2$. Since $V^{\Gamma} = d P_+$, we obtain 
 $E_N(\gamma^V)  = \log (1 + {1\over d})$. It implies: 
 
\be E_D(\gamma^V)\leq {E_N}(\gamma^V) = \log (1 + {1\over d}) < 1 \leq K_D(\gamma^V),\ee
which demonstrates a desired gap between distillable key and distillable entanglement:
\be
E_D(\gamma^V) < K_D(\gamma^V).
\ee

\subsection{Relative entropy of entanglement and \pdits}
In this section we will consider the entanglement contents of the \pbit\ in terms of a measure
of entanglement called relative entropy of entanglement, defined as follows:
\be
E_r(\rho) = \inf_{\sigma_{sep} \in SEP}S(\rho|\sigma_{sep})
\label{eq:relent}
\ee 
where  $S(\rho||\sigma) = -S(\rho) -\tr\rho\log\sigma$ 
is the relative entropy, and $SEP$ is the set of separable states.
In the  Section \ref{sec:relent} we will show, that for {\it any} state, 
the relative entropy of entanglement is an upper bound 
on the key rate, that can be obtained from the state (for generalizations of this 
result to a wide class of entanglement monotones see \cite{EkertCHHOR2006-ABEkey,Christandl-phd}). 
It is then easy to see, that for any \pbit\ $\gamma$, $E_r(\gamma)$ is greater than 
$\log d$ since $K_D(\gamma)\geq \log d$
by definition of \pdits. The question we address here, is the {\it upper} bound on 
the relative entropy of the \pdit. We  relate its value to the states which 
appear on the \side\ of the \pdits, when Alice and Bob get key by measuring
the \main\ of the \pdit. The theorem below states it formally.

{\theorem  For any 
\pdit\  $\gamma_{ABA'B'}\in \bcal({\cal C}^d\ot{\cal C}^d\ot {\cal C}^{d_A'}\ot{\cal C}^{d_B'})$, which
is secure in standard basis, let $\rho_{A'B'}^{(i)}$ 
denote states which appears on \side\ of the \pbit, after obtaining outcome $ii$ 
in measurement performed  in standard basis on its \main.  Then
we have 
\be
 E_r(\gamma_{ABA'B'}) \leq \log d + {1\over d}\sum_{i=0}^{d-1}E_r(\rho_{A'B'}^{(i)}) 
\label{th: pbitrelent}
\ee
where $AB$ denotes key and $A'B'$ \side part of the \pdit.
\label{th:erpbit}
}

{\proof  One can view the quantity ${1\over d}\sum_{i=0}^{d-1}E_r(\rho_{A'B'}^{(i)}) $
as the relative entropy of $\gamma_{ABA'B'}$ dephased on $AB$ in computational basis \cite{lock-ent}. In case $d= 2^k$, it can be easily done with applying unitary $U_i$ - random sequence of $\sigma_z$ and $I$ unitary transformations. In general case one can use the so
called Weyl unitary operators (see e.g. \cite{FukudaHolevo_Weylchan}). Such an implementation of dephasing uses $\log d$ bits of randomness. 

Following the proof of non-lockability of relative entropy of entanglement \cite{lock-ent} (see also  \cite{LPSW1999}), we can write
\be
E_r(\gamma_{ABA'B'}) - E_r(\sum_i p_i \sigma_i) \leq \log d
\ee
where $\sigma_i = U_i \ot I_{A'B'} \gamma_{ABA'B'} U_i^{\dagger} \ot I_{A'B'}$ and $p_i = {1\over d}$.  As we have observed 
above, the relative entropy of dephased state $\sum_i p_i \sigma_i$ equals 
${1\over d}\sum_{i=0}^{d-1}E_r(\rho_{A'B'}^{(i)})$
which ends the proof. 
}

The above theorem is valid also for regularized relative entropy, defined as \cite{DonaldHR2001}
\be
E_r(\rho)^{\infty}=\lim_{n\rightarrow \infty} {1\over n}E_r(\rho^{\ot n}).
\ee

{\theorem Under the assumptions of theorem \ref{th:erpbit} there holds:
\be
 E_r^{\infty}(\gamma_{ABA'B'}) \leq \log d + {1\over d}\sum_{i=0}^{d-1}E_r^{\infty}(\rho_{A'B'}^i),
\ee
\label{th:regrelentpbit}
}

For the proof of this theorem, as rather technical, we refer the reader to Appendix  \ref{app:relentpdit}.

\subsection{Irreducible \pbit\ - a unit of privacy}
In Section  \ref{sec:charact} we have characterized states which contain ideal key i.e. {\pdits}. A {\pdit\ } has an
$AB$ subsystem called here the \main.  $\log d$ bits of key can be obtained from such a {\pdit\ } by a
complete measurement in some basis performed on this \main\ of {\pdit\ }. However, as it follows from the
characterization given in theorem \ref{th:pdit}, {\pdits\ } have also the $A'B'$ subsystem, called here the \side.  
This part can also serve as a source of key. Indeed there are plenty of such {\pdits\ } that contain more than $\log d$ key, 
due to their \side. Therefore not every {\pdit\ } can serve as a of unit of privacy and we need the following
definition:
{\defin Any {\pdit\ } $\gamma $ (with $d$-dimensional \main) for which $K_D(\gamma) = \log d$ is
called irreducible.
}

This definition distinguishes those $\pdits$ for which measuring their \main\ is the optimal protocol for
drawing key. They are called {\it irreducible} in opposite to those, which can be reduced by distillation 
protocol to some other {\pdits\ } which have more than $\log d$ of key. Irreducible {\pdits\ } 
are by definition units of privacy
(although they are not generally interconvertible).

Determining the class of irreducible {\pdits\ } is potentially a difficult task, as it leads to optimisation over protocols 
of key distillation. However we are able to show a subclass of {\pdits}, which are irreducible.
To this end we use a result, which is proven in Section \ref{sec:relent}, namely that the relative entropy of entanglement
is an upper bound on distillable key. Having this we can state the following proposition:
{\prop  Any \pdit\ $\gamma$, with $E_r(\gamma) = \log d$,  is irreducible.
}

{\proof  By definition of $\pdit$ we have $K_D(\gamma) \geq \log d$ and by theorem \ref{erbound}
 from Section  \ref{sec:relent} we have $K_D(\gamma) \leq E_r(\gamma)$ which is in turn less than $\log d$ by
assumption, and the assertion follows.
}
We can provide now a class of  \pdits\ which have $E_r = \log d$ and by the above proposition are irreducible.
These are \pdits\ which have separable states that appear on \side\ conditionally on outcomes of complete measurement
on \main\ part in the computational basis. 
{\prop 
For any \pdit\  $\gamma_{ABA'B'}\in \bcal({\cal C}^d\ot{\cal C}^d\ot {\cal C}^{d_A'}\ot{\cal C}^{d_B'})$, which
is secure in standard basis, if $\rho_{A'B'}^{(i)}$ 
denote states which appear on \side\ of the \pbit, after obtaining outcome $|i\>|i\>$ 
in measurement performed  in standard basis on its \main\ are separable states,  then {\pdit\ } $\gamma_{ABA'B'}$
is irreducible.
}

{\proof  Due to bound on relative entropy of  \pdit\  given in theorem \ref{th:erpbit} we have 
that $E_r(\gamma)$ is less then or equal to $\log d$ since conditional states $\rho_{A'B'}^{(i)}$ 
are separable and hence have relative entropy of entanglement equal to zero. 
$E_r(\gamma)$ is also not less then $\log d$, since it is greater than the amount of distillable
key, which ends the proof.
}
 
Note, that examples (\ref{eq:explpbit}), (\ref{eq:explflow}) given in Section \ref{ssec: pbtexpls} fulfill the assumptions 
of this theorem, and are therefore irreducible \pbits. They are also the first known non trivial states (different than pure state) for which the amount of distillable key has been calculated. Using the bound of relative entropy on distillable key, one can also show, that the class of {\it maximally correlated states} has $K_D = E_D = E_r$, since for the latter $E_D = E_r$.

\section{Approximate {\pbits }} 
\label{sec:apbits}
We present here a special property of states which are close to \pbit. 
We have already seen, that  \pbits\  have similar properties to the maximally entangled EPR states.
In particular, the norm of the upper-right block in standard form
as well as in $X$-form of \pbit\ is equal to ${1\over 2}$. We will show here, 
that for general states the norm of that block tells how close the state 
is to a \pbit: any state which
is close in trace norm to \pbit\ must have the norm of this block close to ${1\over 2}$, and vice versa.

We will need the following lemma that relates 
the value of coherence to the distance from the maximally entangled state for two qubit states.

{\lemma For any bipartite state $\rho_{AB}\in B({\cal C}^2\ot {\cal C}^2)$  
expressed on the form 
 $\rho_{AB} = \sum_{ijkl=0}^{1}a_{ijkl} |ij\>\<kl|$ we have:
\be
\tr \rho_{AB}P_+ \geq 1-\ep \Rightarrow  Re(a_{0011})> {1\over 2}-\ep
\label{lem:trelem}
\ee
and
\be
Re(a_{0011})> {1\over 2}-\ep  \Rightarrow  \tr \rho_{AB}P_+ \geq 1-2\ep
\label{lem:elemtre}
\ee
\label{lem:2qcoh}
}

{\proof  For the proof of this lemma, see Appendix \ref{app:approx-pbits}. }

We can prove now that approximate \pbits\  have norm of an appropriate block 
close to ${1\over 2}$.

{\prop  If the state 
$\sigma_{ABA'B'} \in B({\cal C}^2\ot{\cal C}^2 \ot {\cal C}^d\ot{\cal C}^{d'}) $ 
written in the form $\sigma_{ABA'B'} = \sum_{ijkl=0}^{1} |ij\>\<kl|\ot A_{ijkl}$  fulfills 
\be
|| \sigma_{ABA'B'} - \gamma_{ABA'B'}|| \leq \ep
\ee
for some \pbit\ $\gamma$, then for $0< \ep <1$ there holds $|| A_{0011}|| \geq {1\over 2}-\ep$.
\label{th:beqnorm1}
}

{\proof
The \pbit\ $\gamma$ is a twisted EPR state, 
which means that there exists twisting $U$ which applied to basic \pbit\ $P_+\ot \rho$
gives $\gamma$. 
We apply this $U$ to both
states $\sigma_{ABA'B'}$ and $\gamma$ and trace out the $A'B'$ subsystem
of both of them.
Since these operations can not increase the norm distance 
between these states, so that we have for 
$\sigma_{AB}=\tr_{A'B'}U\sigma_{ABA'B'}U^{\dagger}$
\be
||\sigma_{AB} - P_+ || \leq \ep.
\ee

It implies, by equivalence of  norm and fidelity (\ref{lemfidnor}) that 
\be
F(\sigma_{AB},P_+ ) \geq 1-{1\over 2}\ep.
\ee
We have also that $F(\sigma_{AB},P_+)^2=\tr\sigma_{AB}P_+$ so that 
\be
\tr\sigma_{AB}P_+ >1 - \ep
\ee
for $\ep <1$.  Now  by lemma (\ref{lem:2qcoh}) this yields 
$|a_{0011}|\geq Re(a_{0011})\geq {1\over 2} - \ep$, 
where $a_{0011}$ is coherence of the state $\rho_{AB}=\sum_{ijkl=0}^1 a_{ijkl}|ij\>\<kl|$.
However, we have 
\be
|a_{0011}|=|\tr  U_{00} A_{0011} U_{11}^\dagger|
\ee
where $U_{00}$ and $U_{11}$ come from twisting, that we have applied. 
Using now the fact that $\|A\|=\sup_U \tr AU$, where supremum is 
taken over unitary transformations we get 
\be
\|A_{0011}\|\geq |a_{0011}| \geq 1-\ep.
\ee
This ends the proof. \blacksquare
}

Now we will formulate and prove the converse statement, saying that 
when the norm of the right upper block is close to $1/2$, 
then the state is close to some \pbit.

{\prop
If the state $\sigma_{ABA'B'} \in B({\cal C}^2\ot{\cal C}^2 \ot {\cal C}^d\ot{\cal C}^{d'}) $ 
with a form $\sigma_{ABA'B'} = \sum_{ijkl=0}^{1} |ij\>\<kl|\ot A_{ijkl}$  fulfills 
$|| A_{0011}|| > {1\over 2}-\ep$ then for $0< \ep <{1\over 8}$ there exists \pbit\ $\gamma$ such, that  
\be
|| \sigma_{ABA'B'} - \gamma_{ABA'B'}|| \leq \delta(\ep)
\ee
with $\delta(\ep)$ vanishing, when $\ep$ approaches zero. More specifically,
\be
\delta(\epsilon)= 2\sqrt{8\sqrt{2\ep}+ h(2\sqrt{2\ep})} + 2\sqrt{2\ep}
\ee
with $h(.)$ being the binary entropy function $h(x)=-x\log x -(1-x)\log(1-x)$.
\label{th:beqnorm2} 
}

{\proof  In this proof by $\rho_{X}$ we denote respective reduced density 
matrix of the state $\rho_{ABA'B'}$. 
Let $\rho_{AB}$ be the privacy-squeezed state of the state $\sigma_{ABA'B'}$
i.e. $\rho_{AB} = Tr_{A'B'}\rho_{ABA'B'}$ where $\rho_{ABA'B'}=U_{ps}\sigma_{ABA'B'}U_{ps}^{\dagger}$ for certain twisting $U_{ps}$.
By definition the entry $a_{0011}$ of $\rho_{AB}$ is equal to $||A_{0011}||$. 
By assumption we have, $a_{0011}=||A_{0011}|| > {1\over 2}-\ep$. 
By lemma \ref{lem:2qcoh} (equation (\ref{lem:elemtre})) we have
that 
\be
\tr\rho_{AB}P_+ > 1-2\ep.
\ee 
We have then
\be
F(\rho_{AB},P_+)^2=\tr\rho_{AB}P_+ 
\ee
which, by equivalence of norm and fidelity (\ref{fidnor}) gives
\be
||\rho_{AB} - P_+||\leq 2\sqrt{2\ep}.
\label{b1}
\ee

Let us now consider the state 
$\rho_{ABA'B'}= U_{ps}\sigma_{ABA'B'}U_{ps}^{\dagger}$ and its purification 
to Eve's subsystem $\psi_{ABA'B'E}$ so that we have:
\be
\rho_{AB} = \tr_{A'B'E} (\psi_{ABA'B'E}) 
\label{eq:rosigma}
\ee
By the Fannes inequality (see eq. (\ref{eq:Fannes_ineq}) in Sec. \ref{subsec:notation}) we have
that 
\be
S(\rho_{AB}) = S(\rho_{A'B'E}) \leq 4\sqrt{2\ep} \log d_{AB} + h(2\sqrt{2\ep}).
\label{fannesent}
\ee
>From this we will get that $||\psi_{ABA'B'E} - \rho_{AB}\ot \rho_{A'B'E}||$ is of order of 
$\ep$.  We prove this as follows.
Since norm distance is bounded by relative entropy as follows \cite{OhyaPetz}
\be
{1 \over 2} ||\rho_1 - \rho_2||^2 \leq S(\rho_1|\rho_2),
\label{normerdist}
\ee
one gets:
\be
||\psi_{ABA'B'E} - \rho_{AB}\ot \rho_{A'B'E}|| \nonumber
\leq \sqrt{2 S(\psi_{ABA'B'E}||\rho_{AB}\ot \rho_{A'B'E})}.
\label{lastineq}
\ee
The relative entropy distance of the state to it's subsystems is equal to 
quantum mutual information 
\be
I(\psi_{AB|A'B'E}) = S(\rho_{AB}) + S(\rho_{A'B'E}) - S(\psi_{ABA'B'E}),
\ee
(We will henceforth use shorthand notation $I(X:Y)$, $S(X)$).
which gives
\be
I(AB:A'B'E) = 2S(AB) \leq 2(4\sqrt{2\ep} \log d_{AB} + h(2\sqrt{2\ep})).
\ee
where last inequality comes from Eq. (\ref{fannesent}). Coming back to 
inequality (\ref{lastineq}) we have that 
\be
||\psi_{ABA'B'E} - \rho_{AB}\ot \rho_{A'B'E}|| \leq \sqrt{2 I(AB:A'B'E)} \leq 
2\sqrt{4\sqrt{2\ep} \log d_{AB} + h(2\sqrt{2\ep})}
\ee
If we trace out the subsystem $E$ the inequality is preserved:
\be
||\rho_{ABA'B'} - \rho_{AB}\ot \rho_{A'B'}|| \leq 2\sqrt{8\sqrt{\ep} + h(2\sqrt{\ep})},
\label{b2}
\ee
where we have put $d_{AB}=4$, as we deal with pbits. Now by triangle 
inequality one has:
\be
||\rho_{ABA'B'}-P_+\ot \rho_{A'B'}||\leq ||\rho_{ABA'B'}-\rho_{AB}\ot \rho_{A'B'} ||+
 ||\rho_{AB}\ot \rho_{A'B'} - P_+\ot \rho_{A'B'}||.
\ee
We can apply now the  bounds (\ref{b1}) and (\ref{b2}) to the  above inequality obtaining
\be
||\rho_{ABA'B'}-P_+\ot \rho_{A'B'}||\leq 2\sqrt{8\sqrt{2\ep} + 
h(2\sqrt{2\ep})} + 2\sqrt{2\ep}.
\ee
Let us now apply the twisting $U_{ps}^{\dagger}$ (transformation which is inverse to 
twisting $U_{ps}$)  to both states on left-hand-side of the above  
inequality.  Since $\rho_{ABA'B'}$ is defined as
 $U_{ps}\sigma_{ABA'B'}U_{ps}^{\dagger}$ we get that:
\be
||\sigma_{ABA'B'}-U_{ps}^{\dagger}P_+\ot \rho_{A'B'}U_{ps}||\leq 2\sqrt{8\sqrt{2\ep}+ h(2\sqrt{2\ep})} + 2\sqrt{2\ep},
\ee
i.e. our state is close to \pbit\ $\gamma =U_{ps}^{\dagger}P_+\ot \rho_{A'B'}U_{ps}$.
Then the theorem follows with $\delta(\ep) = 
2\sqrt{8\sqrt{2\ep}+ h(2\sqrt{2\ep})} + 2\sqrt{2\ep}$.} \blacksquare

{\rem The above propositions establish the norm of  upper-right block of matrix 
(written in computational basis according to ABA'B' order of subsystems), as a 
parameter that measures closeness to \pbit, and in this sense 
it measures security of the bit obtained from the \main. 
The state of form (\ref{blockform}) 
is close to a pbit if and only if the norm of this block is close to ${1\over 2}$. 
This is the property of approximate pbits, however
it seems not to have an analogue for approximate pdits with $d \geq 3$.}

\section{Expressing Alice and Bob states in terms of Eve's states}
\label{sec:eve-states}
In this section we will express the state $\rho_{ABA'B'}$ in such a way that 
one explicitly sees Eve's states in it. We will then interpret the results of the
previous sections in terms of such a representation. In particular, we will see 
that  the norm of the upper-right block not only measures closeness to \pbit, but  
it also measures the security of the bit from the \main\, directly,
in terms of fidelity between corresponding Eve's states.  

\subsection{The case without \side. "Abelian" twisting.}
Consider first the easier case of a state without \side\, i.e.
\be
\rho_{AB}=\sum_{ij i'j'} \rho_{iji'j'} |ij\>\<i'j'|.
\ee
Purification of this state is of the following form
\be
\psi_{ABE}=\sum_{ij} \sqrt{p_{ij}} |ij\>_{AB} |\psi_E^{ij}\>
\ee
where $p_{ij}=\rho_{ijij}$.
We see, that when Alice and Bob measure the state in basis $|ij\>$,
Eve's states corresponding to outcomes $ij$ are $\psi_{ij}$,
and they occur with probabilities $p_{ij}$.
Performing partial trace over Eve's system,
one obtains
\be
\rho_{AB}= \sum_{iji'j'} \sqrt{p_{ij}p_{i'j'}} \<\psi^{i'j'}_E|\psi_E^{ij}\>
|ij\>\<i'j'|
\ee
Thus the matrix elements of $\rho_{AB}$ are inner products of Eve's states.
If we have all inner products between set of states,
we have complete knowledge about the set, up to a total unitary rotation,
which is irrelevant for security issues (since Eve can perform this herself).
Thus density matrix $\rho_{AB}$ can be represented in such a way
that all properties of Eve's states are explicitly displayed.
Moreover, moduli of matrix elements are related to fidelity between
Eve's states:
\be
|\rho_{iji'j'}|=\sqrt{p_{ij}p_{i'j'}}\, F(\psi_E^{ij},\psi_E^{i'j'}).
\label{eq:fid1}
\ee

\paragraph{Two-qubit case.}
For example, for two qubits, the density matrix looks as follows
(we have not shown all elements)
\be
\rho_{AB}=
\left[\bea{cccc}
p_{00} & \times &\times & \sqrt{p_{00} p_{11}} \<\psi_E^{11}|\psi_E^{00}\>
\\
\times & p_{01}&\times &\times \\
\times & \times & p_{10} &\times \\
\times & \times & \times &p_{11} \\
\eea\right]
\label{eq:irho-by-eve}
\ee
Let us now consider the conditions for having one bit of perfect key  
obtained from the measurement in the two qubit case. They are as follows:
(i)  $p_{00}=p_{11}=1/2$ and (ii) $\psi_{00}=\psi_{11}$ up to a phase factor. The
latter condition is equivalent to $F(\psi_{00},\psi_{11})=1$ (we have
dropped here the index $E$). The two conditions can
be represented by a single condition:
\be
\sqrt{p_{00}p_{11}}\,F(\psi_{00},\psi_{11})={1\over 2}
\ee
However, we know from (\ref{eq:fid1}) that this means
that upper-right matrix element of $\rho_{AB}$ should satisfy $|\rho_{00
11}|=1/2$.
Consider now approximate bit of key, so that 
the conditions are satisfied up to some accuracy. 
Again we can combine them into single condition 
\be
\sqrt{p_{00} p_{11}} F(\psi_{00}^E,\psi_{11}^E)>{1\over 2} - \ep
\ee
This translates into 
\be
|\rho_{00 11}|\geq {1\over 2} -\epsilon,
\ee
which gives: 
\be
F^2\leq {1\over 2} (1+2|\rho_{0011}|),
\ee
where $\psi_{ME}={1\over \sqrt 2}(e^{i\phi_{00}}|00\>+e^{i\phi_{11}}|11\>)$, and $F^2=\<\psi_{ME}|\rho|\psi_{ME}\>$.

Thus, in particular for $F=1$, we must have
$|\rho_{0011}|=1/2$.
The change of phases can be viewed as a unitary operation,
where phases are controlled by the basis $|ij\>$:
\be
U=\sum_{ij} |ij\>\<ij|e^{i\phi_{ij}}
\label{eq:ab-twist}
\ee
Since we have $\psi_{\max}=U|\psi_+\>$, this operation can be
called "abelian" twisting. Abelian because only phases are
controlled. Thus we can summarize our considerations by the following
statement.
{\it A two-qubit state has perfectly secure one bit of key with respect to basis
$|ij\>$, if and only if it is a twisted EPR state (by abelian twisting of Eq.
(\ref{eq:ab-twist})):
\be
\rho_{AB}=U|\psi_+\>\<\psi_+|U^\dagger.
\ee
Moreover, if a state satisfies security condition approximately,
it must be close in fidelity to some state $U\psi_+$. The quality of the bit of key 
is given by magnitude of a c-number $|\rho_{0011}|$}.
\subsection{The general case.}
In this section we will represent in terms of Eve's states the state 
which has both \main\ and \side. We will see then, how the twisting becomes
"nonabelian",
and the condition of closeness to pure state $U\psi_+$ changes into that
of closeness to \pbit. If we write state in basis of system
$AB$
(\main) we get blocks $A_{iji'j'}$ instead of matrix elements
\be
\rho_{ABA'B'}=\sum_{ij i'j'} |ij\>_{AB}\<i'j'| \ot A^{iji'j'}_{A'B'}.
\label{eq:state_block_form}
\ee
After suitable transformations (see Appendix \ref{subsec:App_Eves_states} for details), we arrive
at the following form:
\be
\rho_{ABA'B'}=\sum_{iji'j'} \sqrt{p_{ij}p_{i'j'}}|ij\>_{AB}\<i'j'|\ot
\bigl (\ucal_{ij} \sqrt{\rho_E^{ij}}
\sqrt{\rho_E^{i'j'}}\ucal_{i'j'}^\dagger \bigr )^T,
\ee
where the operator $\ucal_{ij}^\dagger\equiv U_{ij}\wcal V_{ij}$ maps the space
$\hcal_{A'B'}$
exactly onto a support of $\rho_E^{i'j'}$ in space $\hcal_E$, and the dual
operator
$\ucal_{ij}=V_{ij}^\dagger \wcal^\dagger U_{ij}^\dagger$
maps the support of $\rho_E^{ij}$ back to $\hcal_{A'B'}$.


Let us note, that in parallel to Eq. (\ref{eq:fid1}) we have that the {\it
trace norms}
of the blocks $A$ are connected with fidelities between Eve's states
\be
\|A_{iji'j'}\|=\sqrt{p_{ij}p_{i'j'}}\,F(\rho_E^{ij},\rho_E^{i'j'})
\ee

\paragraph{The case of two qubit \main}

If the \main\ is two qubit system we get
\be
\rho_{ABA'B'}=
\left[\bea{cccc}
p_{00}[\ucal_{00} \rho_E^{00}\ucal_{00}^\dagger ]^T & \times
&\times & \sqrt{p_{00} p_{11}}\, [\ucal_{00} \sqrt{\rho_E^{00}}\,
\sqrt{\rho_E^{11}}\,\ucal_{11}^\dagger ]^T \\
\times & p_{01}[\ucal_{01} \rho_E^{01}\ucal_{01}^\dagger ]^T &\times
&\times \\
\times & \times &p_{10}[\ucal_{10} \rho_E^{10}\ucal_{10}^\dagger ]^T
&\times \\
\times & \times & \times & p_{00}[\ucal_{11}
\rho_E^{11}\ucal_{11}^\dagger ]^T\\
\eea\right]
\label{eq:rho-by-eve}
\ee
Let us now discuss conditions  for presence of one bit of key. They are again
(i) $p_{00}+p_{11}={1\over 2}$ and (ii) Eve's states are the same
$\rho_E^{00}=\rho_E^{11}$. This is equivalent to
\be
\sqrt{p_{00}p_{11}}\, F(\rho_E^{00},\rho_E^{11})={1\over 2}
\ee
which is nothing but trace norm of upper-right block $\|A_{0011}\|$.
Also conditions for approximate bit of key requires the norm
to be close to ${1\over 2}$.
Moreover, to see how \pbit\ and the twisting arise,
let us put all Eve's states equal to each other,
and probabilities corresponding to perfect correlations. We then obtain
\be
\rho_{ABA'B'}={1\over d}\sum_{ij} |ii\>_{AB}\<jj|\ot
[\ucal_{ii} \rho_E \ucal_{jj}^\dagger ]^T.
\ee
where $\rho_E$ is one fixed state, that Eve has irrespectively of
outcomes. We see here almost the form of \pbit. One difference might be
is that instead of usual unitaries,
we have some embeddings $\ucal_{ii}$. However, since now Eve's space
is of the same dimension as $A'B'$ (because Eve has single state),
they are actually usual unitaries. The transposition does not really make a
difference,
as it can be absorbed both by state, and by unitaries.
It is interesting to see here in place of phases from previous
section the unitaries appeared, so that abelian twisting changed into
nonabelian one.
Also the condition for key changed from modulus of c-number - matrix
element, to a trace norm of q-number -  a block.

\section{Overview}
\label{sec:overview}
In this section we will shortly summarize what we have done so far. Then
we will describe the goals of the paper, and briefly outline how we will achieve
them.
\subsection{{\Pbits\ } and twisting}
We have considered a state shared by Alice and Bob,
which was divided into two parts: the \main\ $AB$ and the \side\
$A'B'$.
The \main\ is measured in a local basis, while the \side\ is kept. The
latter
is seen by Eve as an environment that may restrict her knowledge
about
outcomes of measurement performed on the pdit.

We have shown two important facts. First, we have characterized all the states
for which measurement on the \main\ gives  perfect key. The states are called
\pdits, and they have a very simple form.
Moreover, we have shown that twisting does not change the \ccq state
arising from measurement on the \main\ part. (We should emphasize here,
that twisting must be controlled by just the same basis
in which the measurement is performed.)

This is an interesting feature, because twisting may be a nonlocal
transformation.
Thus even though we apply a nonlocal transformation to the state,
the quality of the key established by measuring the \main\
(in the same basis) does not change.
>From the exhibited examples of \pbits, we have seen that some of them have
very small distillable entanglement. Since \pbits\  are EPR states subjected to
twisting, we see that in this case the twisting must have been
very nonlocal, since it significantly diminished
distillable entanglement.
Because \pbits\ contain at least one bit of secure key,
we have already seen that distillable key can be much larger than
distillable entanglement. 

However our main goal is to show that
there are {\it bound entangled} states from which one can draw key. Thus we need
distillable entanglement to be strictly zero.
Here it is easily seen that any perfect \pdit\ 
is an NPT state. Even more, one can show that \pdits\  are always distillable 
\cite{AH-pditdist}.
Thus we cannot realize our goals by analysing perfect \pbits.

\subsection{Approximating \pbits\  with PPT states}
After realising that \pbits\ cannot be bound entangled,
one finds that this still does not exclude bound entangled
states with private key. Namely, even though bound entangled states
cannot contain exact key (as they would be \pbits\  then)
they may contain {\it almost exact} key. Such states
would be in some sense close to \pbits.
Note that this would be impossible, if the only states containing perfect
key were maximally entangled state. Indeed, for $d\ot d$ system 
if only a state has greater overlap than $1/d$ with a maximally entangled state
we can distill singlets from it \cite{BBPS1996}.

Recall, that for a state with \main\ being two qubits,
the measure of quality of the bit of key coming from measuring
the \main\ is trace norm of upper-right block. The key is perfect 
if the norm is $1/2$ (we have then \pbit) and
it is close to perfect, if the norm is close to $1/2$.
Thus our first goal will be to find bound entangled states
having the trace norm of that block arbitrarily close to $1/2$.
We will actually construct such PPT states (hence bound entangled) in
Sections \ref{subsec:new-family},
\ref{subsec:approx}.
In this way we will show that there exist bound entangled
states that contain an arbitrarily (though not perfectly) secure bit of key.
\subsection{Nonzero rate of key from bound entangled states}
It is not enough to construct bound entangled states with arbitrary secure single bit of
key.
The next important step is to show that given many copies
of BE states one can draw nonzero asymptotic rate of secure key.
To show this we will employ (in Section \ref{subsec:secure-from-be}) the BE
states
with almost perfect bit of key.
Let us outline here the most direct way of proving the claim. 

To be more specific, we will
consider many copies of states $\rho_\epsilon$
which have upper-right block trace norm equal to $1/2-\epsilon$.
We will argue that one can get
key by measuring the \main\ of each of them,
and then process via local classical manipulations and public discussion
the
outcomes. How to see that one can get nonzero rate in this way?

We will first argue, that the situation is the same,
as if the outcomes were obtained from a state which is close to maximally entangled.
To this end we will apply the idea of \cohsq\ 
described in Section \ref{ssec:twistcosq}.

First recall, that we have shown that operation of twisting
does not change security of \ccq state - more precisely,
it does not change the state of the Eve's system and \main\
of Alice and Bob systems, which would arise, if Alice and Bob measured 
the \main. Thus whatever twisting we will apply,
from cryptographic point of view the situation will not change.
The total state will change,
yet this can be noticed only by those who have access to
the \side\ of Alice and Bob systems, and Eve does not have such access.

We will choose such a twisting, that will change the upper-right block of $\rho^\ep$ 
into a positive operator.
This is exactly the one which realizes \cohsq\  of this state. 
Now, even though security is not changed,
the state is changed in a very favorable way for our purposes.
Namely, we can now trace out the \side,
and the remaining state of the \main\ (a \cosqd\ state of the initial one) will be close to maximally entangled.
Indeed, twisting does not change trace norm of the upper-right block.
Because now the block is a positive operator,
its trace norm is equal to its trace, and tracing out \side\
amounts just to evaluating trace of blocks.
Since the trace norm was $1/2 -\epsilon$, the upper-right element
of the state of \main\ is is equal to $1/2 -\epsilon$,
which means that state is close to maximally entangled (where the
corresponding element is equal to $1/2$).
One can worry, that it is now not guaranteed that
the security is the same, because we have performed not only twisting,
but also partial trace over \side.
However the latter operation could only make situation worse,
since partial trace means giving the traced system to Eve. 

Now the only remaining thing is to
show that we can draw key from data
obtained by measuring many copies of state close to an EPR state,
then definitely we can draw key from many copies of more secure \ccq state
obtained from our $\rho^\epsilon$.
To achieve the goal, we thus need some results about drawing
key from \ccq state. Let us recall that we work in scenario,
where Alice and Bob are promised to share i.i.d. state, 
so that the ccq state is tensor product of identical copies. In this case, the  
needed results have been provided in
\cite{DevetakWinter-hash}. It follows, that the rate of key is at least
$I(A:B)-I(A:E)$, where $I$ is mutual information.
If instead of almost-EPR state, we have just an EPR state,
the above quantity is equal to $1$. Indeed, perfect correlations,
and perfect randomness of outcomes gives $I(A:B)=1$ and purity of the EPR state
gives $I(A:E)=0$. Since we have state close to an EPR state,
due to continuity of entropies, we will get  $I(A:B)\approx 1$, and
$I(A:E)\approx 0$.
Thus given $n$ copies of states that approximate \pbits\  one can
get almost $n$ bits of key in limit of large $n$.

\subsection{Drawing key and transforming into \pbits\  by LOCC}
Apart from showing that key can be drawn from BE states, we want to
develop the theory of key distillation from quantum states.
To this end in Section \ref{sec:def-rate-key} we recast definition of
distilling key in terms of distilling \pbits\  by local operations
and classical communication.
This is important change of viewpoint: drawing key requires referring to
Eve; while distilling \pbits\  by LOCC concerns solely bipartite
states shared by Alice and Bob, and never requires explicit referring
to Eve's system. Thus we are able to pass from the 
game involving three parties: Alice, Bob and Eve
to the two players game, involving only Alice and Bob.

We will employ two basic tools: (i) the concept of making a protocol coherent;
(ii) the fact (which we will prove) that having almost perfectly secure \ccq
state
is equivalent to having a state close to some \pdit.
Note that in one direction, the reasoning is very simple:
if we can get nonzero rate of asymptotically perfect \pdits\  by LOCC,
we can also measure them at the end, and get in this way
asymptotically perfect \ccq states, which is ensured by item (ii) above.
The converse direction is a little bit more involved: we take any protocol
that produces key, apply it coherently, and this gives pure final state
of Alice, Bob and Eve's systems. From (ii) it follows,
that the total state of Alice and Bob must be close to \pdit.

Let us briefly discuss how we will show the fact (ii).
The essential observation is that both the \ccq state  $\tilde \rho_{ABE}$ and
Alice and Bob total state $\rho_{ABA'B'}$
are reductions of the same pure state $\psi_{ABA'B'E}$. Here some explanation 
is needed: in general, since the \ccq state is obtained 
by measurement, it is not reduction of  $\psi_{ABA'B'E}$. However, one can first 
apply measurement  coherently to the state $\rho_{ABA'B'}$.  
Then the \ccq of the new state 
$\rho_{ABA'B'}'$ is indeed the reduction of $\psi_{ABA'B'E}'$. In the actual 
proof we will proceed in a slightly different way.

Now, if we have two nearby \ccq states, we can find their purifications 
that are close to each other too. Then also Alice and Bob states
arising when we trace out Eve's system are close to each other (because
partial trace can only make states closer).
The whole argument is slightly more complicated,
but the above reasoning is the main tool.

The equivalence we obtain 
puts
the task of drawing key into the standard picture of
state manipulations by means of LOCC. The theory of such manipulations
is well developed and, in particular, there are quite general
methods of obtaining bounds on transition rates (in our case the transition
rate is
just distillable key), see \cite{Michal2001}. Indeed, we will be able to
show that
relative entropy of entanglement is an upper bound for distillable key. The
main
idea of deriving the bound is similar to the methods from LOCC state
manipulations. However significant obstacles arise, to overcome which we
have developed essentially new tools. 

\section{Two definitions of distillable key: LOCC and LOPC paradigms}
\label{sec:def-rate-key}
In this section we show that distillable amount of \pdits\  by use of LOCC 
denoted by $K_D $ is equal to classical secure key distillable by means 
of {\it local operations and public communication} (LOPC).  

\subsection{Distillation of \pdits\ }
\label{subsec:pdit-rate}
We have established a family of states - \pdits\  - which have the following 
property: after measurement in some basis $\bcal$ they give a perfect dit of 
key.  In entanglement theory one of the important aims is to distill singlets 
(maximally entangled states) which leads to
operational measure of distillable entanglement. We will pose now 
an analogous task namely distilling \pdits\ (private states) which are of the 
form (\ref{eq:ipdit}). This gives rise to a definition of distillable key i.e. 
maximal achievable rate of distillation of pdits.  Similarly as 
in the case of distillation of singlet, it is usually 
not possible to distill exact pdits. Therefore  the formal definition of 
distillable key $K_D$  will be  a bit more involved.

{\defin For any given state 
$\rho_{AB}\in \bcal(\hcal_A\otimes \hcal_B)$  
let us consider sequence $P_n$ of
$LOCC$ operations such that $P_n(\rho_{AB}^{\ot n})=\sigma_n$, where
$\sigma_n \in B(\hcal^{(n)}_A\ot\hcal^{(n)}_B)$. 
A set of operations ${\cal P} \equiv \cup_{n=1}^{\infty} \{P_n\}$ is called
\pdit\  distillation protocol of state $\rho_{AB}$ if there holds
\be
\lim_{n\rightarrow \infty} ||\sigma_n-\gamma_{d_n}|| = 0,
\ee
where $\gamma_{d_n}$ 
is a \pdit\ whose  key  part is of dimension $d_n\times d_n$.

For given protocol $\cal P$, its rate is given by
\be
{\cal R}({\cal P})=\limsup_{n\rightarrow \infty} {\log d_n \over n}
\ee
The distillable key of state $\rho_{AB}$ is given by 
\be
K_D(\rho_{AB})=\sup_{\cal P}{\cal R}(\cal P).
\ee
\label{kd}
}
In other words, due to this definition, Alice and Bob given $n$ copies of state
$\rho_{AB}$ try to get a state which is close to some pdit state
with $d=d_n$. 
Unlike so far in entanglement theory, effect of distillation of quantum key 
depends not only on the number $n$ of copies of initial state  
but also on the choice of the output state. 
This is because private dits appears not to be reversibly 
transformable with each other by means of LOCC operations, as it is 
in case of maximally entangled states in LOCC entanglement distillation.
Thus the quantity $\kd$ is a rate of distillation to the large 
class of states.  (Of course, since the definition involves optimization,
$K_D$ is well defined; in particular the expensive \pdits\ will be suppressed).

One can be interested now if this new parameter of states $K_D(\rho)$ 
has an operational meaning for quantum cryptography. 
One connection is obvious: given a quantum state 
Alice and Bob may try to distill some pdit state, 
and hence get (according to the above definition) 
$\kd(\rho_{AB})$ bits of key if such distillation has
nonzero rate. However the question arises: 
is it the best way of extraction of a classical secure key from a quantum state?
I.e. given a quantum state is the  largest amount of classical key 
distillable from a state equal to $K_D$. 
We will give to this question a positive answer now. It means, that distilling
private dits i.e. states of the form (\ref{eq:pdit}) is the best way of distilling classical
key from a quantum state.

\subsection{Distillable classical secure key: LOPC paradigm}
\label{subsec:def-cd}
The issue of drawing classical secure key from a quantum state is formally 
quite different from the definition of drawing \pdits. However 
it will turn out that it is essentially the same thing. 
In the LOCC paradigm, we have an initial state $\rho$ hold by Alice and 
Bob  who apply to it an LOCC map, and obtain a final state $\rho'$.
Thus the LOCC paradigm is essentially a {\it bipartite} paradigm. 

In the paradigm of drawing secure classical key (see e.g. \cite{Christandl:2002,EkertCHHOR2006-ABEkey}),  there are three parties, Alice, Bob and Eve. 
They start with some joint state $\rho_{ A  B E}$ 
where subsystems $ A, B,E$ belong to Alice, Bob and Eve respectively. 
Now, Alice and Bob essentially perform again some LOCC operations.
However we have now a tripartite system, 
and we should know how that operation act on the whole system. 







{\defin An operation $\Lambda $ belongs to LOPC class if it is composition of 
\bei
\item[(i)] Local Alice (Bob) operations, i.e. operations of the form 
$\Lambda_A\ot I_{BE}$ (or $\Lambda_B\ot I_{AE}$).
\item[(ii)] Public communication from Alice to Bob (and from Bob to Alice).  
E.g. the process of communication  from Alice to Bob is described 
by the following map 
\be
\Lambda(\rho_{aABE})= \sum_i P_i \rho_{aABE} P_i \ot |i\>_b\<i|\ot |i\>_e\<i|
\ee
where $P_i=I_{ABE}\ot |i\>_a\<i|$. Here the subsystem $a$ carries the message to be sent,
and the subsystems $b$ and $e$ of Bob and Eve represent the received message.
\eei
\label{def:lopc}
}

Let us note, that LOCC operations can be defined in the same way, 
the only difference is that we drop Eve's systems (both $E$ and $e$) and 
corresponding operators. 


Now, drawing secure key  means  obtaining the following state 
\be
\rhoideal =\sum_{i=0}^{d-1}{1\over d} |ii\>\<ii|_{AB} \otimes \rho^E
\label{eq:ideal-ccq}
\ee
by means of LOPC. Since output states usually can not be exactly $\rho^{ccq}_{ideal}$ Alice and Bob
will get state  of the \ccq form (\ref{eq:ccq-def}) i.e. 
\be
\rhoreal=\sum_{i,j=1}^{d}p_{ij}|ij\>_{AB}\<ij|\otimes\rho_{ij}^E
\label{eq:real-ccq}
\ee
There are two issues here:
first, Alice and Bob should have almost perfect correlations, 
second, Eve states should have small correlations  with states $|ij\>$ of Alice 
and Bob systems. The first condition refers to  {\it uniformity},
the second one to {\it security}.
There are several ways of quantifying these correlations, and some of them are 
equivalent. To quantify security  \cite{Ben-OrHLMO05} one  can use Holevo function of 
distilled \ccq state, namely:
\be
\chi(\rho_{ccq})\equiv S(\rho_E) - 
\sum_{i,j=1}^{d}p_{ij}S(\rho_{ij}) \leq \ep
\label{eq:sec-holevo}
\ee
where $S(\rho)=\tr\rho\log\rho$ denotes von Neumann entropy, and  
\be
\rho_E=\sum_{i,j=1}^{d}p_{ij}\rho_{ij}.
\ee
Alternatively, one can use similar condition based on norm
\be
\sum_{ij} p_{ij} \|\rho_E - \rho^E_{ij}\|\leq \ep
\label{eq:sec-norm}
\ee
The condition of maximal correlations between Alice and Bob (uniformity) can be   
of the following form
\be
||\sum_{i,j=1}^{d}p_{ij}|ij\>\<ij| -{1\over d}\sum_{i=1}^{d}|ii\>\<ii| || \leq \ep
\label{eq:unif}
\ee
 One can also use again the trace norm between the real state (\ref{eq:real-ccq})
that is obtained and the ideal desired state (\ref{eq:ideal-ccq}) as done 
in \cite{DevetakWinter-hash},
which includes both maximal correlations condition as well as security condition.
The condition says that the state $\rhoreal$ obtained by Alice and Bob 
is closed to some ideal state
\be
\|\rhoreal - \rhoideal\|\leq \ep
\label{eq:joint}
\ee
We will discuss relations between this condition,
and security criteria (\ref{eq:sec-holevo}) and (\ref{eq:sec-norm})
as well as with uniformity criterion (\ref{eq:unif})
in Appendix \ref{sec:securitycrit} . 

For the purpose of definition of secret key rate in this paper, we 
apply the joint criterion (\ref{eq:joint}).
Consequently, we adopt the following measure of distillable classical secure key  from 
a quantum tripartite state:

{\defin For any given state 
$\rho_{ABE}\in B(\hcal_A\ot \hcal_B\ot\hcal_C)$  
let us consider sequence $P_n$ of
$LOPC$ protocols such that $P_n(\rho_{ABE}^{\ot n})=\beta_{n}'$, where
$\beta'_n$ is \ccq state 
\be
\beta_{n}'=\sum_{i,j=0}^{d_n-1}p_{ij} |ij\>\<ij|_{AB}\otimes\rho^E_{ij}
\label{genccq}
\ee
from 
$\bcal(\hcal^{(n)}) =B(\hcal^{(n)}_{A}\ot\hcal^{(n)}_{B}\ot\hcal^{(n)}_E)$ 
with $\dim \hcal^{(n)}_A = \dim \hcal^{(n)}_{B}= d_n$.
A set of operations ${\cal P} \equiv \cup_{n=1}^{\infty} \{P_n\}$ is called
classical key distillation protocol of state $\rho_{AB}$ if there holds
\be
\lim_{n\rightarrow \infty} ||\beta_{n}'-\beta_{d_n}|| = 0,
\ee
where $\beta_{d_n} \in B(\hcal^{(n)}$ is of the form
\be
{1\over d_n}\bigl(\sum_{i=1}^{d_n} |ii\>_{AB}\<ii|\bigr)\ot \rho^E_n,
\ee
$\rho^E_n$ are arbitrary states from $B(\hcal_{E}^{(n)})$. The rate of a protocol $\pcal$ is given by 
\be
{\cal R}({\cal P})=\limsup_{n\rightarrow \infty} {\log d_n \over n}
\ee
Then the distillable classical key of state $\rho_{ABE}$ is defined as 
supremum of rates
\be
C_D(\rho_{ABE})=\sup_{\cal P}{\cal R}(\cal P).
\ee
\label{ckd0}
}

The above definition works for any input tripartite state $\rho_{ABE}$. 
However in this paper we are only interested in the case 
where the total state is pure. The latter is  determined 
by state $\rho_{AB}=\tr_E \rho_{ABE}$ up to unitary transformations 
on Eve's side. Since from the very definition $C_D$ does not change 
under such  transformations,  the latter freedom is not an issue, 
so that we can say the state $\rho_{AB}$  completely determines 
the total state. Thus we  get definition of distillable classical 
secure key from {\it bipartite } state $\rho_{AB}$ 
{\defin
For given bipartite state $\rho_{AB}$ the distillable 
classical secure key is given by  
\be
C_D(\rho_{AB})\equiv  C_D(\psi_{ABE})
\ee
where $\psi_{ABE}$  purification of $\rho_{AB}$.
\label{def:ckd}
}

\subsection{Comparison of paradigms}

Let us compare two definitions \ref{kd} and \ref{def:ckd} 
of distilling cryptographical key. The difference 
is mostly that the first one deals only with bipartite system,
and the goal is to get the desired final state by applying a class 
of LOCC operations. 
Within the second paradigm, we have tripartite state 
and we want to get a wanted state by means of LOPC operations. 
Thus the first paradigm is much more standard in quantum information 
theory. The second one comes from classical security theory
(see e.g. \cite{MaurerWolf00CK}), where probability distributions of triples 
of random variables $P(X,Y,Z)$ are being processed.


In the next section we will see that if the tripartite initial 
state is pure, the two paradigms are tightly connected. 
In the case of distillation of exact key, they are almost obviously identical, 
while in the inexact case, the only issue is to make 
the asymptotic security requirements equivalent.
We will see that an output \pdit\  obtained by LOCC 
implies some \ccq state obtained by LOPC, and vice versa.

\subsection{Composability issues}
In the present paper we consider the promised scenario, 
which is the first step to consider in unconditionally secure QKD. 
\footnote{The next step was subsequently done in papers \cite{uncond-PRL,uncond-ieee}}. 
The latter security definition is required to be {\it universally composable}, which means that 
a QKD protocol can be used as a subroutine of any other cryptographic protocol
\cite{Ben-Or-Mayers-compos,Ben-OrHLMO05,RennerK04-key}. 
To this end, one needs to choose carefully the measure of security. 
Our starting point is the LOPC paradigm, where the measure of security 
is the trace norm. This is compatible with \cite{Ben-OrHLMO05}, where 
it was shown that such security measure implies indeed composability (see eq. (10) of  \cite{Ben-OrHLMO05}). In particular, if we concatenate $n$ QKD protocols with 
security $\ep$ measured by trace norm  between the obtained state and the 
ideal target, the overall protocol will have security bounded by $n\ep$. 
Thus trace norm in LOPC paradigm can be called {\it composable security condition}.

However, we want to recast QKD within the LOCC paradigm, i.e. in terms of  distance between 
Alice and Bob  states rather than tripartite states. In this spirit, 
in \cite{Ben-OrHLMO05} it is shown  that fidelity with maximally entangled state 
implies composability, in the sense that 
if fidelity is $1-\ep$ then the norm is bounded by $\sqrt \ep$ (see eq. (20) of \cite{Ben-OrHLMO05}). 
This may seem a bit uncomfortable, because while composing many protocols
we have now to add not epsilons,  but rather their square roots. However 
one can see that it is in general not possible to do better while starting from fidelity. 
Due to inequality $1-F(\sigma,\rho)\leq {1\over 2}||\rho-\sigma||$ 
\cite{Fuchs-Graaf} we can equally well use the trace norm. Indeed it will give us at worse
 the $\sqrt{\ep}$ bound for composable security condition in the LOPC paradigm.

In our paper we have a more general situation, i.e. 
we deal with distance to private states rather than just maximally entangled states. 
According to the above discussion, we have decided to 
use trace norm in LOCC scenario. In our techniques, in the mid-steps 
we shall use fidelity, hence we are again left with $\sqrt\ep$. 
We do not know, whether one can omit fidelity, and get rid of the square root.

Having said all that, we should emphasize that finally 
the minimal requirement is that security is an exponentially decreasing 
function of some security parameter, whose role can play e.g. the number 
of all qubits in the game. This condition is of course not spoiled by square root.  
(Example of a situation, where  this requirement is not met was conjectured  
in \cite{Ben-OrHLMO05} and then proved  in \cite{KoenigRBM-locking}). 
In our paper we shall derive equivalences between various security conditions,
and the word ``equivalence'' will be understood in the sense of not spoiling 
the proper exponential dependence on total number of qubits. In particular, the security
conditions will be called equivalent also when they differ by the factor polynomial
in number of qubits (see e.g. \cite{Gottesman-Lo}).


\section{Equality of key rates in LOCC and LOPC paradigms}
\label{sec:equality}
In this section we will show that definitions 
\ref{kd} and \ref{def:ckd} give rise to the same quantities.
In this way  the problem of drawing key within original LOPC 
paradigm is recast in terms of transition to a desired state 
by LOCC.  First we will describe a {\it coherent version} of LOPC protocol.
Then we will use it to derive equivalence in exact case (where 
protocols produce as outputs ideal \ccq states 
or ideal \pdits). Subsequently we will turn to the general case 
where inexact transitions are allowed.

\subsection{Coherent version of LOPC key distillation protocol}
\label{coh}
The main difference between LOPC and LOCC paradigms 
is that in  the first one we have transformations between tripartite states shared by Alice, Bob and Eve, while in the latter one - between bipartite states  shared by Alice and Bob.
Thus in LOPC paradigm, the part of the state held by Alice and Bob does not, in general tell
us about security. To judge if Alice and Bob have secure key we  need the whole 
$\rho_{ABE}$ state.  Security is assured by the lack of correlations of this state with Eve.   
Thus if we want to recast the task of drawing key in terms of LOCC paradigm, we need to consider
such LOCC protocol which produces output state that assure security of key {\it itself}. 
We will do this by considering coherent version of LOPC key distillation protocols 
(cf. \cite{balance,DevetakWinter-hash}). 

The most important feature of the version will be that 
given any LOPC protocol, starting with some initial  pure state $\psi_{A B E}$ 
and ending up with some \ccq state $\rho_{ABE}$, 
its coherent version will end up with a state $\psi'_{AA'BB'E}$ such 
that tracing out $A'B'$ part will give exactly the \ccq state $\rho_{ABE}$. 
In this way, the total Alice and Bob state $\rho_{AA'BB'}$ 
will keep the whole information about Eve (because up to unitary on Eve's system,
purification is unique).

In  {\it coherent} version of key distillation protocol 
Alice and Bob perform their local operations  in a coherent way 
i.e. by {\it adding ancillas}, performing {\it unitary transformations} and 
{\it putting aside} appropriate
parts of the system. This additional part of the system is  
discarded in the usual protocol. However, holding 
this part allows one to keep the total state of Alice Bob and Eve
pure in each step of the protocol.  This is because we use pure
ancillas, pure initial state and apply only unitary transformations which
preserve purity.

Alice and Bob can also perform public communication. Its coherent version
is that e.g. Bob and Eve apply C-NOT operations to Alice's subsystem which holds the result of measurement.

Formally, the coherent version of process of communication from definition \ref{def:lopc} 
is an operation of the form
\be
\Lambda(\rho_{aABE})=U(\rho_{aABE}\ot |0\>_{a'}\<0|\ot|0\>_b\<0|\ot|0\>_e\<0|)U^\dagger
\ee
where 
\be
U=I_{ABE}\ot \sum_i|i\>_a\<i|\ot U_{a'}^{(i)} \ot U_b^{(i)}\ot U_e^{(i)}
\ee
with unitary transformation $U_e^{(i)}$,$U_b^{(i)}$ satisfying
$U_e^{(i)}|0\>_e=|i\>_e$ and similarly for $U_b^{(i)}$, $U_{a'}^{(i)}$. To finalize the operation, Alice puts aside 
the system $a'$. 

Such a coherent version of LOPC protocol has the following two features:
(i) keeps the state pure.
(ii) after tracing out subsystems that are put aside 
we obtain exactly the same state as in the original protocol.

Now we are in position to construct for a given LOPC protocol 
a suitable LOCC protocol, which will output \pbits\ when the former 
protocol will output ideal key. Namely, the local operations from LOPC 
protocol, we replace with their coherent versions. The public 
communication we take in original - incoherent - form (of course, without broadcast to Eve since we now deal with bipartite states).
One notes that such operation produces the same state of Alice and Bob 
as the one produced by coherent version of the original LOPC protocol traced out 
over Eve's system.  This is because, if we trace out Eve, then there is no difference 
between coherent and incoherent versions of public communication. 

Thus from an LOPC protocol we have obtained some LOCC protocol - a special one, 
where local systems are not traced out. In this way one gets  a bridge which joins the 
two approaches, and shows that different definitions of distillable key
are equivalent. In particular, suppose that the LOPC  protocol produced the ideal \ccq state.
Then the output of LOCC protocol obtained as a coherent version of  this protocol
will produce a state which (due to theorem \ref{th:pdit}) must be a \pdit.

{\rem Of course, the notion of coherent version need not concern just some LOPC 
protocol. Also an LOCC operation, 
that contained measurements and partial traces can be made coherent,
which in view of the above considerations means simply, that 
the systems are not traced out, but only  "kept aside"
and  measurements are replaced by appropriate local unitaries. 
Actually, if we include all pure ancillas that will be 
added in the course of realizing the LOCC operation,  
the coherent version of the operation is nothing but 
a  {\it closed} LOCC operation  \cite{huge-delta} introduced 
for sake of counting local resources such as local information. 
One can think of the shield part, as being the state of all the lab equipment
and quantum states, left over from the process of key distillation.  
}

\subsection{Equivalence of paradigms: The case of exact key}
Here we will consider the ideal case, where the distillation of the key gives {\it exactly} 
the demanded output state. One can state it formally and observe:

{\prop Let $K_D^{exact}$ and $C_D^{exact}$ denote optimal rates 
achievable by LOPC and LOCC protocols which as outputs have 
exact \ccq states (\ref{eq:ideal-ccq}) and \pdit\ states (\ref{eq:pdit}),
respectively. Then for any state $\rho_{AB}$ we have 
\be
K_D^{exact}(\rho_{AB}) = C_D^{exact}(\rho_{AB})
\ee}


{\proof 
If after LOPC protocol $\pcal$, Alice and Bob obtained exact $d\times d$ \ccq state 
(\ref{eq:ideal-ccq}), then the coherent application of $\pcal$ 
due to theorem \ref{th:pdit} and discussion of section \ref{coh} will produce \pdit\ of the same dimension.  Conversely,  if by LOCC Alice and Bob can get  a \pdit, 
then after measurement, again by theorem \ref{th:pdit}, they will obtain 
exact \ccq state (\ref{eq:ideal-ccq}) of the same dimension.  \blacksquare
%
}

\subsection{Distillation of classical key and distillation of pdits - equivalence in 
general (asymptotically exact) case}

We will prove here the theorem,  which implies, that even in 
nonexact case, distillation of 
\pdits\  from initial bipartite state 
by LOCC is equivalent to distillation of key by LOPC 
from initial pure state, that is purification of the bipartite state. 
This in turn means that the rates in both paradigms are equal. 

{\theorem
Let Alice and Bob share a state $\rho$ such that Eve has it's purification. 
Then the following holds: if Alice and Bob can distill by LOPC operations 
a state such that with Eve's subsystem it is \ccq state i.e. 
of the form
\be
\rho_{ABE}=\sum_{i,j=1}^{d} p_{ij} |ij\>\<ij|_{AB}\otimes\rho_{ij}^E,
\label{lem:rowkey}
\ee
with $||\rho_{ABE}^{ccq} - \rho_{ideal}^{ccq}|| \leq \ep$, 
then they can distill by LOCC operations a state $\rho_{out}$ 
which is close to some pdit state $\gamma$ in trace norm:
\be
||\rho_{out} - \gamma||\leq \sqrt{2\ep},
\ee
where the \main\ of a pdit $\gamma$ is of dimension $d\times d$.\\
Conversely, if by LOCC they can get state $\rho_{out}$ satisfying 
$||\rho_{out} - \gamma||\leq \ep$, then by LOPC they can get state $\rho_{ccq}$
satisfying $||\rho_{ABE}^{ccq} - \rho_{ideal}^{ccq}|| \leq \sqrt{2\ep}$
\label{equivlemma2}
}

{\proof 
The "if" part of this theorem is proven as follows. By assumption Alice and Bob 
are able to get by some LOPC protocol $\pcal$ a \ccq\ state  $\rho_{ABE}$ satisfying
\be
||\rho_{ABE}^{ccq} - \rho_{ideal}^{ccq} ||\leq \ep.
\ee
Now by equivalence  between norm and fidelity (Eq. (\ref{fidnor}) of Appendix) we 
can rewrite this inequality as follows
\be
F(\rho_{ABE}^{ccq},\rho_{ideal}^{ccq})>1-{1\over 2}\ep.
\ee
By definition of fidelity
\be
F(\rho,\sigma)=\max_{\psi,\phi}|\<\psi|\phi\>|
\ee
where maximum is taken  over all purifications $\psi$ and $\phi$ of $\rho$ and $\sigma$ 
respectively, we can fix one of these purification arbitrarily, and 
optimise over the other one.  Let us then choose such a purification
$\psi_{ABA'B'E}$ of $\rho_{ABE}^{ccq}$ which is the output of coherent application 
of the mentioned protocol $\pcal$. There exists purification $\phi_{ABA'B'E}$ of $\rho_{ideal}$ 
such that it's overlap with $\psi$ is greater than $1-{1\over 2}\ep$. 
Since the fidelity can only increase after partial trace applied to both the states, it will be still greater than 
$1-{1\over 2}\ep$ once we trace over Eve's subsystem. 
Thus we have 
\be
F(\rho_{ABA'B'}^{\psi},\sigma_{ABA'B'}^{\phi})>1-{1\over 2}\ep.
\ee
where $\sigma_{ABA'B'}^{\phi}$ and $\rho_{ABA'B'}^{\psi}$  
are partial traces of $\phi$ and $\psi$ respectively. 
The state $\sigma_{ABA'B'}^{\phi}$ (partial trace of $\phi$) comes from purification 
of an ideal state, and by 
the very definition it is some pdit state $\gamma$. At the same time, the state $\rho_{ABA'B'}^{\psi}$ (partial trace
of $\psi$) is the one which is the output of coherent application of protocol $\pcal$. 
Thus by coherent version of $\pcal$ Alice and Bob can obtain state 
close to \pdit\ which proves the "if" part of the theorem.

To obtain equivalence let us prove now the converse implication. 
The proof is a sort of "symmetric reflection" of the proof of 
the previous part. 

This time we assume that there exists LOCC protocol starting with $\rho$, ending
up with final state $\rho_{out}$ with 
\main\ od $d\times d$ dimension which is close to some  \pdit\ in norm i.e. 
\be
||\rho_{out}-\gamma|| \leq \ep.
\ee 
Due to equivalence between fidelity and norm, we have 
\be
F(\rho_{out},\gamma)\geq 1-\ep/2
\ee
The total state after protocol is $\psi_{ABA'B'E}$,
and if partially traced over Eve it returns $\rho_{out}$. Then we can find 
such $\phi$, purification of $\gamma$, that $F(\psi,\phi)>1-\ep$. 
Now let Alice and Bob measure the \main\ and trace out the \side. 
Then out of $\psi$ we get some \ccq state $\rho^{ccq}_{out}$.
The same operation applied to $\phi$ gives ideal ccq state (\ref{eq:ideal-ccq}) 
$\rho_{ideal}^{ccq}$. The operation can only increase the fidelity, so that 
\be
F(\rho_{out}^{ccq},\rho_{ideal}^{ccq})\geq 1-\ep/2
\ee
Returning to norms we get 
\be
||\rho_{out}^{ccq}-\rho_{ideal}^{ccq}||\leq \sqrt{2 \ep}.
\ee
}\blacksquare

\section{Distilling key from bound entangled states}
\label{sec:bound}
In this section we will provide a family of states. Then we will show that 
for certain regions of parameters they have positive partial transpose (which  means that they are non-distillable). 
Subsequently, we shall show that out of the above PPT states 
one can produce, by an LOCC operation, states arbitrarily close to \pbits\ (which also implies that they are
entangled, hence bound entangled). More 
precisely, for any $\ep$ we will find PPT states,
from which by a LOCC protocol, one gets with some 
probability a state $\ep$-close to some \pbit. Since LOCC preserves the PPT property, 
this shows that \pbits\ can be approximated with arbitrary accuracy by PPT states, in sharp contrast with maximally entangled states. 
We then show  how to get from a state  sufficiently close to a \pbit\ 
with non vanishing asymptotic rate of key. We obtain it by reducing the problem 
to drawing key from states that are close to the maximally entangled state.

\subsection{The new family of PPT states ...}
\label{subsec:new-family}
Here we will present a family of states, and will determine the range of parameters 
for which the states are PPT. 
The idea of construction of the family is based on the so called {\it hiding states}
found by Eggeling and Werner in \cite{WernerHide}. Let us briefly recall this result. 
In  \cite{hiding-prl,hiding-ieee} it was shown that one can  
hide one bit of information in two states by correlating the bit of 
information with a pair of 
states  which are almost indistinguishable by use of LOCC operations, yet being 
almost distinguishable by global operations. 
The resulting state with the hidden bit is of the form:
\be
\rho_{hb} = {1\over 2} |0\>\<0|_{AB} \otimes \rho^{1}_{hiding} + {1\over 2}|1\>\<1| 
\otimes \rho^{2}_{hiding}
\ee
In \cite{WernerHide} it was shown that there are {\it separable} states, 
which can serve as arbitrarily good hiding states.
These states are
\be
\tau_1 = ({\rho_s +\rho_a \over 2})^{\otimes k}, \quad \tau_2 = ({\rho_s})^{\otimes k},
\label{hidings}
\ee
where $\rho_s$ and $\rho_a$ are symmetric and antisymmetric Werner states (\ref{eq:Werner-sym-asym}). The higher is the parameter $k$, the
more indistinguishable by LOCC protocols the states become. 

We adopt the idea of hiding bits  to hide entanglement.
Namely instead of bits one can correlate two orthogonal maximally entangled states with these two 
hiding states and get the state:
\be
\rho_{he} = {1\over 2} |\psi_+\>\<\psi_+|_{AB} \otimes \tau_1^{A'B'} + {1\over 2}|\psi_-\>\<\psi_-|_{AB} 
\otimes \tau_2^{A'B'}
\label{he}
\ee 

Let us recall, that our purpose is  to get 
the family of states which though entangled are not distillable, and can approximate 
pdit states. Then the choice of $\rho_{he}$ as a starting point has double advantage.  
First, because $\tau_1$ and $\tau_2$ are hiding, $\rho_{he}$  
will not allow for distillation of entanglement by just distinguishing them. 
Second, the hiding states are separable, so they do not bring in any entanglement 
to the state $\rho_{he}$.
However the state (\ref{he}) is obviously NPT.  Indeed, consider 
partial transposition of $BB'$ system. It is  composition of partial 
transpositions  of $B$ and $B'$ subsystems. 
If one applies it to the state (\ref{he}), one gets
\ben
&&\rho_{ABA'B'}^\Gamma=(I_A\ot T_B\ot I_{A'}\ot T_{B'})(\rho_{ABA'B'})= \nonumber \\ 
&&=\left[\bea{cccc}
{1\over 2}({\tau_1+\tau_2\over 2})^{\Gamma} &0&0&0 \\
0& 0&{1\over 2}{({\tau_1-\tau_2\over 2})}^{\Gamma}&0 \\
0&{1\over 2}({{\tau_1-\tau_2}\over 2})^{\Gamma}&0& 0\\
0 &0&0& {1\over 2}({\tau_1+\tau_2\over 2})^{\Gamma}  \\
\eea
\right]\nonumber\\
\een
where ${\Gamma}$ denotes partial transposition over subsystem $B'$ (as partial transposition over B caused interchange
of blocks of matrix of (\ref{he})). This matrix is obviously not positive for the lack of middle-diagonal blocks. To prevent
this we admix a separable state ${1\over 2}(|01\>\<01| + |10\>\<10| )\otimes \tau_2$ with a probability $(1-2p)$, where $p \in {(0,{1\over 2}]}$. It's matrix reads then
\be
\rho_{(p,d,k)}=\left[\bea{cccc}
p({\tau_1+\tau_2\over 2}) &0&0&p({\tau_1-\tau_2\over 2}) \\
0& ({1\over 2}-p)\tau_2&0&0 \\
0&0&({1\over 2}-p)\tau_2& 0\\
p{({\tau_1+\tau_2\over 2})} &0&0& p({\tau_1+\tau_2\over 2})  \\
\eea
\right],
\label{row-key}
\ee
In subscript we explicitly write the parameters on which this state depends implicitly: $d= d_{A'}=d_{B'}$ is the dimension of
symmetric and antisymmetric Werner states used for hiding states (\ref{hidings}) and $k$ is parameter
of tensoring in their construction.  We shall see, that for some range of $p$, almost every state of this family is a PPT state. 
We formalise it in the next lemma. 

{\lemma   Let  $\rho_a\in B({\cal C}^d\otimes{\cal C}^d) $ and $\rho_b\in B({\cal C}^d\otimes{\cal C}^d)$ be symmetric
and antisymmetric Werner states respectively, and let  $k$ be such that
\be
\tau_1 = ({\rho_s +\rho_a \over 2})^{\otimes k}, \quad \tau_2 = ({\rho_s})^{\otimes k}
\ee
holds. Then for any $p \in (0, {1\over 3}]$ and any $k$ there exists $d$ such that 
state (\ref{row-key}) has positive partial transposition.
More specifically, the state (\ref{row-key}) is PPT if and only if 
the following conditions are fulfilled
\ben
&&0<p\leq{1\over 3}\\ \nonumber
&&{1-p\over p}\geq \biggl({d\over d-1}\biggr)^k 
\een
\label{lem:pptfam}
}

For the proof of this lemma see Appendix \ref{subsec:App_proof_lem:pptfam}.

\subsection{... can approximate pdits}
\label{subsec:approx}
We have just established a family $\rho_{(p,d,k)}$ such that for certain $p$, 
$k$ and $d$ they are PPT states. We will then show, that 
by LOCC one can transform some of them  to a state 
close to \pbits. More precisely, for any fixed accuracy,
we will always find $p$, $d$ and $k$ such that 
it is possible to reach \pbit\ up to this accuracy,
starting from some number of copies of $\rho_{(p,d,k)}$ and applying LOCC
operations. 

Subsequently, we will show that one can always choose 
the initial states $\rho_{(p,d,k)}$ to be PPT. 
Since LOCC operations do not change PPT property,
we will in this way show that there are PPT states
that approximate \pbits\  to arbitrarily high accuracy. 


We will first prove the following theorem. 

{\theorem  For any $\ep >0$ and any  $p \in ({1\over 4},1]$ 
there exist state $\rho$ from family of state $\{\rho_{(p,d,k)}\}$ 
(\ref{row-key}) such that for some $m$ from $\rho^{\ot m}$ 
one can get  by LOCC (with nonzero probability of  success) a state 
$\sigma$ satisfying $|| \sigma - \gamma || \leq \ep$ for 
some private bit $\gamma$.
\label{closetopbit}
}

{\proof 
First of all let us notice that by theorem \ref{th:beqnorm2} it is enough to show, 
that one can transform $\rho^{\ot m}$ into a state $\rho'$ which has sufficiently 
large norm of the upper-right block.

Let Alice and Bob share $m$ copies of a state 
$\rho$ from the family (\ref{row-key}). The number $m$ and parameters $(p,k,d)$ of this 
state will be fixed later.  

Now let Alice and Bob apply the well known recurrence protocol - ingredient of 
protocols of distillation of singlet states \cite{BDSW1996}. Namely they take one system in
state $\rho$ as {\it source system}, and iterate the following procedure.  In $i$-th step they take one system in state $\rho$, and treat it as a {\it target system}. Let us remind that both systems have four subsystems $A$, $B$, $A'$ and $B'$. To distinguish the source and target system, the corresponding subsystems of a target system we call $\tilde{A},\tilde{B},\tilde{A'},\tilde{B'}$. On the source and target system they both perform a CNOT gate with a source at the $A(B)$ part of a source system and target at $\tilde{A}(\tilde{B})$ part of a target system for Alice (Bob) respectively. Then, they both measure the $\tilde{A}$ and $\tilde{B}$ subsystem of the target system in computational basis respectively, and compare the results. If the results agree, they proceed the protocol, getting rid of the $\tilde{A}\tilde{B}$ subsystem. If they do not agree, they abort the protocol. With nonzero probability of success they can perform this operation $m-1$ times having
each time the same source system, and some fresh target system in state $\rho$. That is they start with $m$ systems in state $\rho$ and in each step (upon success) they use up one system and pass to the next step. 



One can easily check, that the submatrices (blocks)  of the 
state $\rho_{(p,d,k)}^{rec}$ which survives $m-1$ steps 
of this recurrence protocol (which clearly happens with nonzero probability) are equal to the $m$-fold tensor power of the 
elements of initial matrix $\rho_{(p,d,k)}$:
\be
\rho^{rec}_{(p,d,k)}={1\over N}\left[\bea{cccc}
[p({\tau_1+\tau_2\over 2})]^{\ot m} &0&0&[p({\tau_1-\tau_2\over 2})]^{\ot m} \\
0& [({1\over 2}-p)\tau_2]^{\ot m}&0&0 \\
0&0&[({1\over 2}-p)\tau_2]^{\ot m}& 0\\
{[p({\tau_1-\tau_2\over 2})]}^{\ot m}&0&0&{[p({\tau_1+\tau_2\over 2})]}^{\ot m}\\
\eea
\right].
\label{eq:rec-state}
\ee
where the normalisation is given by 
\be 
N=\tr[\rho^{rec}_{(p,d,k)}] =  2 p^m +2 ({1\over 2}-p)^m.
\label{trac}
\ee
Let us consider the upper-right block $\tilde A_{0011}$ of the matrix (\ref{eq:rec-state}) without normalisation.
Norm of this block is equal to
\ben
\|\tilde A_{0011}\|=\left({p\over 2}\right)^m||  
({{\rho_a -\rho_s}\over 2})^{\ot k} - {\rho_s}^{\ot k} ||=
\nonumber \\ \left({p\over 2}\right)^m \big(2(1 - 2^{-k})\big)^m
= p^m(1- 2^{-k})^m.
\label{normel}
\een
where second equality is consequence of the fact, 
that $\rho_a$ and $\rho_s$ have orthogonal supports which gives that
$\rho_s^{\ot k}$ is orthogonal to any term in expansion 
of $({{\rho_a -\rho_s}\over 2})^{\ot k}$ but the one ${1\over 2^k}
\rho_s^{\ot k}$. Thus the result is equal to norm 
of $[({{\rho_a - \rho_s}\over 2})^{\ot k} - {1\over 2^k}\rho_s^{\ot k}]$ (which
is $(1-{1\over 2^k})$) plus norm of the 
difference $|{1\over 2^k}\rho_s^{\ot k} - \rho_s^{\ot k}|$ 
which gives the above formula. Thus the norm of the upper-right block $A_{0011}$ of 
the  state (\ref{eq:rec-state}) is given by 
\be
\|A_{0011}\|={1\over N}\| \tilde A_{0011}\| ={1\over 2}(1- {1\over 2^k})^m {1\over 1 + ({1 -2p\over 2p})^m}.
\label{aboveterm}
\ee
We want now to see, if we can make the norm to be arbitrary close to $1/2$.
(then by Lemma \ref{th:beqnorm2} the state will be arbitrary close to a \pbit).
Since $p > {1\over 4}$, we get that  $({1 -2p\over 2p})^m$ converges 
to 0 with $m$. Although increasing $m$ 
diminishes the term $(1- {1\over 2^k})^m $, we can first fix $k$ large enough, so 
that the whole expression (\ref{aboveterm})  will be as close
to ${1\over 2}$ as it is required. } \blacksquare

Now we have the following situation. We know that for $p\in({1\over 4},1]$ if  
${1\over 2}(1- {1\over 2^k})^m {1\over 1 + ({1 -2p\over 2p})^m}$ 
is close to  $1/2$, then the state (\ref{eq:rec-state}) is close 
to \pbit. On the other hand, from lemma (\ref{lem:pptfam}) it follows that 
for (i) $p\in(0,1/3]$ and (ii) ${1-p\over p} \geq \big({d\over d-1}\big)^k$ 
the state $\rho_{(p,d,k)}$ is PPT, 
hence also the state (\ref{eq:rec-state}) is PPT (because it was obtained 
from the former one by LOCC operation). If we now fix $p$ from 
interval $(1/4,1/3]$, then by choosing high $m$ and for 
such $m$, high enough $k$,  then the state (\ref{eq:rec-state})
is close  to \pbit. Now, we can fix also $m$ and $k$, and choose $d$ so large 
that the condition (ii) is also fulfilled so that the state becomes PPT. 
This proves the following theorem, which is main result of this section.

\begin{theorem}
PPT states can be arbitrarily close to \pbit\  in trace norm.
\end{theorem}

Here might be the appropriate place to note an amusing property of state (\ref{he}):  Namely,
Eve knows one bit of information about Alice and Bob's state -- she knows the phase of their Bell state.  But she only has one
bit of information about their state, thus it cannot be that she also knows the bit of their state, which
is the key.  In some sense, giving Eve the bit of phase information, means that she cannot know the bit value.

\subsection{Distillation of secure key}
\label{subsec:secure-from-be}
In the previous subsection we have shown that private bits can be approximated by 
PPT states.  Now, the question
is whether given many copies of one of such PPT states Alice and Bob can 
get nonzero rate of classical key. Below,  we will
give the positive answer.

The main idea of the proof is to show that 
from the PPT state which is close to \pbit\ Alice and 
Bob by measuring, can obtain \ccq state satisfying 
conditions of protocol (DW) found by Devetak and 
Winter \cite{DevetakWinter-hash}. Namely, they have shown  that for an initial 
cqq state (state which is classical only on Alice side; this includes ccq state 
as special case)  between Alice, Bob and Eve, 
\be
C_D(\rho_{ABE}) \geq  I(A:B)-I(A:E)
\label{eq:hash}
\ee 
Here $I(A:B)$ stands for  the quantum mutual information of the state $\rho$ with 
subsystems $A$ and $B$ given by
\be
I(A:B)= S(A) + S(B) - S(AB),
\ee
where $S(X)$ stands for the von Neumann entropy of $X$ (sub)system of the state $\rho$.

Using the above result we can prove now,
that from many copies of states close to a \pbit,
one can draw nonzero asymptotic rate of key. 
\begin{lemma}
If a state $\rho$ is close enough to \pbit\ in trace norm, then 
$K_D(\rho)>0$. 
\end{lemma}

{\proof
The idea of the proof is as follows. Suppose that $\sigma$ is 
close to \pbit\ $\gamma$. We then consider twisting that changes 
$\gamma$ into basic \pbit\ ${P_+}_{AB}\otimes \sigma_{A'B'}$ with some state $\sigma$ on $A'B'$. 
We apply twisting to both states, so that they are still close to 
each other.  Of course, this is only a mathematical tool: 
Alice and Bob cannot apply twisting, which is usually a nonlocal operation. 
The main point is that after twisting, according to theorem \ref{th:ccq} the 
\ccq state does not change.
If we now  trace out systems  $A'B'$ the resulting state 
will be close to maximally entangled, and the  resulting \ccq state  -- at most worse
from Alice and Bob point of view
(because tracing out means giving to Eve). What we have done is just  privacy squeezing of the state $\rho$.
Now, the latter \ccq state has come from measurement of 
a state close to the maximally entangled one. Thus the task reduces  to estimate 
quantities $I(A:B)$ and $I(A:E)$ for a ccq state obtained from measuring the maximally entangled state. 
However due to suitable continuities, first one is close to $1$ 
and second one close to $0$. 
Now by DW protocol, one can draw a pretty high 
rate of key from such \ccq state.
Let us now proceed with the formal proof.

We assume that for some \pbit\ $\gamma$ we have 
\be
||\rho - \gamma||\leq \ep.
\ee
Let us consider twisting $U$ which changes pdit $\gamma$ into a basic pdit. 
Existence of such $U$ is assured by theorem \ref{th:pdit}. If
both states $\gamma$ and $\rho$ are subjected
to this transformation, the norm is preserved, so that 
\be
||U \rho U^\dagger - U\gamma U^\dagger||\leq \ep.
\ee
Also due to  theorem \ref{th:ccq} the \ccq state obtained by measuring 
\main\ of $U\rho U^\dagger$ is the same as that from $\rho$.
Now, the amount of key drawn from such ccq state will not increase 
if we trace out \side. Thus we apply such partial trace to 
$U\rho U^\dagger$ and to $\gamma$, and by monotonicity of trace norm get 
\be
||\tilde \rho_{AB} - P^+_{AB}||\leq \ep.
\label{eq:bliski-singlet}
\ee
It is now enough to show that from \ccq state 
obtained  by measuring $\tilde \rho_{AB}$ (where Eve holds 
the rest of its purification) one can get nonzero rate of key.

To this end let us note, 
that for ccq state obtained from any bipartite state $\rho_{AB}$ by measuring its purification on subsystems A and B in computational basis, we have the following bound for $I(A:E)$:
\be
I(A:E)_{\rm ccq}\leq S(\rho_{AB}).
\ee
Now, since our state $\tilde\rho_{AB}$ is close to $P_+$,
for which $S(P_+)=0$ and $I(A:B)=1$, we can use continuity
of entropy, to bound these quantities  for the state. 
>From Fannes inequality (see eq. \ref{eq:Fannes_ineq} in Sec. \ref{subsec:notation}), we get 
\ben
&&I(A:B)_{\rm ccq} \geq  1 - 4\ep\log d_{AB} - h(\ep),
\label{eq:imbound1} \\ 
&&I(A:E)_{\rm ccq} \leq S(\tilde\rho_{AB}) \leq 8\ep-h(\ep).
\label{eq:imbound2}
\een
To get first estimate, it is enough to note, that 
due to monotonicity of trace norm, 
the estimate (\ref{eq:bliski-singlet}) is also valid 
if we dephase the state $\tilde \rho_{AB}$ 
and $P^+_{AB}$. 
Thus we obtain that 
\be
K_D(\rho)\geq I(A:B)-I(A:E) \geq 1 - 16\ep
\ee
This ends the proof of the lemma.
\blacksquare

Let us note here, that it was not necessary to know,
that the state $\rho$ is close to \pbit. Rather, It was enough to 
know that trace norm of upper right block is close to $1/2$, as we have 
proved that it is equivalent to previous condition (see Sec. \ref{sec:apbits}).
>From this it follows, that after twisting, and tracing out $A'B'$ 
the resulting state $\tilde\rho$ is close to the EPR state, which 
ensures nonzero rate of key (actually the rate is close to 
$1$).

We now can combine the lemma with the fact that we know PPT states 
that are close to $\pbit$, to obtain that there exist PPT 
states from which one can draw secure key. 
The states must be entangled, as from separable states 
one cannot draw key. 
Namely separable state can be established by public discussion. If it could 
then serve as a source
of secret key, one could obtain secret key by public discussion which can not be
possible. For formal arguments see \cite{Curty04:key-ent}. Thus our PPT 
states are entangled. But, since they are PPT,
one cannot distill singlets from them \cite{bound}, hence they are 
bound entangled. In this way we have obtained the following theorem

\begin{theorem}
There exist bound entangled states with $K_D>0$. 
\end{theorem}

We have split the way towards bound entangled states 
with nonzero key into two parts.  First, we have shown that  from 
PPT states $\rho_{(p,d,k)}$ 
by recurrence one can get a state that is close to \pbit. 
Then we have shown, that from a state close to \pbit\ one can draw 
private key. 

Note that we have two quite different steps: recurrence 
was the quantum operation preformed on quantum  Alice and Bob states,
while Devetak-Winter protocol in our case, is classical processing 
of the outputs of measurement. We could unify the picture in two ways. 
First Alice and Bob could measure the \main\ of the initial 
state $\rho_{(p,d,k)}$, and preform recurrence classically 
(since the quantum recurrence is merely coherent application of 
classical protocol). Then the whole process of drawing key from 
 $\rho_{(p,d,k)}$ would be classical (of course, taking into account  
that Eve has quantum states).  
On the other hand, the DW protocol could be applied coheretly, 
so that till the very end, we would have quantum state 
of Alice and Bob.

The result we have obtained allows to distinguish two measures of entanglement 
{\corrolary
Distillable entanglement and distillable classical secure key are different
measures of entanglement i.e. there are states for which there holds
\be
K_D(\rho)>D(\rho)=0.
\ee
}
In further section we will also show that $K_D$ is different than entanglement cost,
as it is bounded by relative entropy of entanglement. 

\section{Relative entropy of entanglement as  upper bound on distillable key}
\label{sec:relent}
In this section we will provide complete proof 
of the theorem announced \cite{pptkey} which
gives general upper bound on distillable key $K_D$. This upper bound is
given by regularised relative entropy of entanglement (\ref{eq:relent}). 
The relative entropy of entanglement \cite{VPRK1997,PlenioVedral1998} is given by 
\be
E_r(\rho)=\inf_{\sigma_{sep}} S(\rho|\sigma_{sep}),
\ee
where $S(\rho|\sigma)=\tr \rho\log \rho - \tr \rho\log \sigma$ is relative entropy, 
and infimum is taken over all separable states  $\sigma_{sep}$. 
The regularized version of $E_r$ is given  by 
\be
E_r^\infty(\rho)=\lim_n {E_r(\rho^{\ot n})\over n}.
\ee
The limit exists, and due to subadditivity of $E_r$,
we have 
\be
E_r^\infty(\rho)\leq E_r.
\ee
It follows that also relative entropy of entanglement is upper bound for $K_D$.

We recall now the following lemma obtained in \cite{pptkey}, the proof of which we provide in Appendix \ref{Appendix:proof_farfromsinglet}: 

{\lemma  Consider a set ${\cal S}^{\tau} := \{U\rho_{ABA'B'}U^{\dagger}\,|\, \rho_{ABA'B'} \in SEP \,,\rho_{ABA'B'} \in B({\cal C}^d\otimes
{\cal C}^d\otimes{\cal C}^{d_{A'}}\otimes{\cal C}^{d_{B'}})\}$
where $U$ is $\bcal$-twisting with $\bcal$ being a standard product basis in $\ccal^d\ot\ccal^d$. Let $\sigma_{ABA'B'}\in {\cal S}^{\tau}$ and $\sigma_{AB} = \tr_{A'B'}\sigma_{ABA'B'}$. We have then 
\be
S(P_+|\sigma_{AB})\geq \log d,
\ee
where $P_+=|\psi^+_d\>\<\psi^+_d|$. 
\label{farfromsinglet}
}

We will also need asymptotic continuity of the 
relative entropy distance from some set of states obtained in \cite{DonaldHR2001} in the form of \cite{Synak05-asym}. 

{\prop  For any compact, convex set of state $\cal S$ that contains maximally mixed state, 
the relative entropy distance from this set given by
\be
E_r^{\cal S} = \inf_{\sigma \in \cal S} S(\rho||\sigma),
\ee
is asymptotically continuous i.e. it satisfies
\be
|E_r^{\cal S}(\rho_1) - E_r^{\cal S}(\rho_2)|< 4 \ep  \log d + h(\ep)
\ee
for any states $\rho_1$, $\rho_2$ acting on Hilbert space $\hcal$ 
of dimension $d$, with $\ep=\|\rho_1-\rho_2\|$ with $\ep\leq 1$.
\label{relatcont}
}

Let us mention, that the original relative entropy of entanglement \cite{PlenioVedral1998}
has in place of $\cal S$ the set of separable states. Another version
has been considered in \cite{Rains2001}, where $\scal$ was set of PPT states.
The latter set has entangled states, but they can be only weakly entangled.
In contrast we will have set in which there may be quite strongly entangled 
states.  

We are now in  position to formulate and prove the main result of this section. 

{\theorem  For any bipartite state 
$\rho_{AB}\in {B}({\cal C}^{d_A}\otimes {\cal C}^{d_B})$ there holds 
\be
K_D(\rho_{AB})\leq E_r^{\infty}(\rho_{AB}),
\ee
\label{erbound}
}

{\proof 
By definition of $K_D(\rho_{AB})$ there exists protocol 
(i.e. sequence of maps $\Lambda_n$),
such that 
\be
\Lambda_n(\rho^{\ot n})=\gamma'_d
\ee
where 
\be
\lim_n {\log d\over n} = K_D(\rho_{AB})
\label{eq:kd-rate}
\ee
and 
\be
\lim_n \|\gamma_d'-\gamma_d\|\equiv  \lim_n\ep_n=0
\label{eq:gammasClose}
\ee
with $\gamma_d$ being \pdit\ with dimension $d^2$ of the key part.

We will present now the chain of (in)equalities, and comment it below.
\ben
&&S(\rho_{AB}^{\ot n}|\tilde{\sigma}_{sep}) \geq S(\gamma_{ABA'B'}'|\sigma_{sep}) =  \label{1ineq}\\
& =&S(U_{\gamma}\gamma_{ABA'B'}'U_{\gamma}^{\dagger}|U_{\gamma}\sigma_{sep}
U_{\gamma}^{\dagger}) \geq  \label{2ineq}\\
& \geq &S(\tr_{A'B'}[U_{\gamma}\gamma_{ABA'B'}'U_{\gamma}^{\dagger}]|\tr_{A'B'}
[U_{\gamma}\sigma_{sep}U_{\gamma}^{\dagger}]) \label{3ineq}\\
&\equiv& S(P_+'|\sigma) \geq  \label{4ineq}\\
& \geq &inf_{\sigma \in T} S(P_+'|\sigma):= E_r^{T}(P_+')\geq \label{5ineq}\\ \label{6ineq}
&\geq & E_r^{T}(P_+) - 4 ||P_+ -P_+'||\log d - h(||P_+ - P_+'||) \geq \\
& \geq &(1- 4 \ep_n)\log d - h(\ep_n) 
\een
Inequality (\ref{1ineq}) is due to the fact, that relative entropy  
does not increase under completely positive maps; in particular
it can not increase under LOCC action applied to it's both arguments
(second argument  becomes other separable 
state since LOCC operations can not create entanglement). In the next 
step, Eq. (\ref{2ineq})  we perform twisting $U_{\gamma}$ controlled 
by the basis in which state $\gamma_m$ is secure (without loss of 
generality we can assume it is standard basis). The equality follows 
from the fact that unitary transformation doesn't change the  relative entropy.
Next (\ref{3ineq}) we trace out $A'B'$ subsystem of both states which only decreases 
the relative entropy. After this operation,
the first argument is $P'_+$, which is a  state close to the EPR state $P_+$. 
($P'_+$ would be equal to the EPR state if $\gamma_{ABA'B'}'$ were 
exactly \pdit)  while second argument  becomes some -- not necessarily 
separable -- state $\sigma$. The state belongs to the set $T$ constructed as follows. 
We take set of separable states on system $ABA'B'$ subject to twisting $U_{\gamma}$ 
and subsequently  trace out  the $A'B'$ subsystem. The inequality (\ref{4ineq}) holds, 
because we take infimum over all states from set $T$ of the function $S(P'_+|\sigma)$. 
This minimised version is named  there $E_r^{T}(P_+')$ as it is relative entropy 
distance of $P'_+$ from the set $T$.

Let us check now, that set $T$ fulfills the conditions of 
proposition \ref{relatcont}. Convexity of this set is obvious, since 
(for fixed unitary $U_{\gamma}$) by linearity it is due to 
convexity of the set of separable states. This set contains the identity
state, since it contains maximally mixed separable state which is unitarily invariant (i.e. invariant under $U_{\gamma}$ ) and 
whose  subsystem $AB$ by definition is the maximally mixed state as well. Thus by 
proposition \ref{relatcont} we have 
that $E_r^{T}$ is asymptotically continuous 
\be
|E_r^{ T}(P_+') - E_r^{ T}(P^+_d)|< ||P_+' - P_+||4 \log d + h(||P_+' - P_+||),
\label{eq:asscont}
\ee
where we assume that the EPR state $P_+$ is of local dimension $d$.
Since $P_+'$ and $P_+$ come out of $\gamma_{AB}'$ and $\gamma_d$ by the same 
transformation described above  (twisting, and partial trace) which doesn't increase 
norm distance, by (\ref{eq:gammasClose}) we have that 
$||P_+' - P_+||\leq \ep_n$. This, together with asymptotic 
continuity (\ref{eq:asscont}) implies (\ref{5ineq}). Now by 
lemma \ref{farfromsinglet} we have 
\be
E_r^{T}(P_+) \geq \log d,
\ee
which gives the last inequality:
\be
E_r^{T}(P_+') \geq (1- 4\ep_n) \log d  - h(\ep_n).
\ee
Summarizing this chain of inequalities (\ref{1ineq})-(\ref{6ineq}), 
we have that for any separable state $\tilde{\sigma}_{sep}$:
\be
S(\rho_{AB}^{\ot n}||\tilde{\sigma}_{sep}) \geq (1- 4\ep_n) \log d  - h(\ep_n)
\ee
Taking now infimum over all separable states $\tilde{\sigma}_{sep}$ we get
\be
E_r(\rho_{AB}^{\ot n}) \geq (1-4\ep_n)\log d - h(\ep_n).
\ee
Now we divide both sides by $n$ and take the limit. 
Then the left-hand-side converges to $E_r^\infty$.
 Due to (\ref{eq:gammasClose}) $\ep_n\to 0$ and due to (\ref{eq:kd-rate}),
$\log d/n \to K_D(\rho_{AB})$. Thus due to continuity of $h$ 
we obtain 
\be
E_r^\infty\geq K_D
\ee
\blacksquare

As an application of the above upper bound, we consider now the relation between distillable key
and entanglement cost. For maximally entangled states these two quantities are of course equal, unlike for general pdits. As an example let us consider again a flower state given in eq. (\ref{eq:explflow}). As follows from \cite{lock-ent}, the flower state has $E_C$ strictly greater
than the relative entropy of entanglement. Since by the above theorem we have
$K_D(\gamma_{flower}) \leq E_r(\gamma_{flower})$, havig $E_r(\gamma_{flower}) < E_C(\gamma_{flower})$, we obtain in this case $K_D(\gamma_{flower})< E_C(\gamma_{flower})$.

Let us note, that the entanglement monotone approach initiated here was then used in full extent in \cite{EkertCHHOR2006-ABEkey,Christandl-phd}. It is shown there, that in fact any bipartite monotone $E$, which is continuous and normalized on private states (i.e. $E(\gamma_d)\geq \log d$), is an upper bound on distillable key. In particular it is shown that the squashed entanglement \cite{Tucci2002-squashed,Winter-squashed-ent} is also an upper bound on distillable key.

\section{A candidate for NPT bound entanglement}
\label{sec:nptbound}

Thus far, all known bound entangled states have positive partial transpose (are PPT).  
A long-standing and interesting open question is whether there exist bound entangled states which are also NPT.
If such states existed, it would imply that the quantum channel capacity is non-additive. 
Since any NPT state is distillable with the aid of some PPT state \cite{wolf}, we would have the curious
property that one can have two states which are each non-distillable, but if you have both states,
then the joint state would be distillable.

We now present a candidate for NPT bound entangled states which are based on the states of
equation (\ref{he}), 
\begin{equation}
\rho_{he} = {1\over 2} |\psi_+\>\<\psi_+|_{AB} \otimes \tau_1^{A'B'} + {1\over 2}|\psi_-\>\<\psi_-|_{AB} \otimes \tau_2^{A'B'} 
\end{equation} 
and which intuitively appear to be bound entangled.  Globally, the flags $\tau_i^{A'B'}$, are distinguishable, but under LOCC
the flags appear almost identical, thus after Alice and Bob attempt to distinguish the flags, the state on $AB$ will be very close to
an equal mixture of $\psi_+$ and $\psi_-$.  The equal mixture of only two different EPR states is separable in dimension $2\times 2$, 
but it is at the edge of separability.
A slight biasing of the mixture, causes the state to be entangled.  Thus, if Alice and Bob are able to obtain even a small amount
of information about which $\tau_i$ they have, they will have a distillable state.  More explicitly, if Alice and Bob attempt distillation
by first guessing 
which hiding state
flag they have, and then grouping the remaining parts of the states into two sets depending on their guess of the hiding state,
they will be left with states of the form
\begin{equation}
\rho_{he} = ({1\over 2}+\epsilon) |\psi_+\>\<\psi_+|_{AB} + ({1\over 2}-\epsilon)|\psi_-\>\<\psi_-|_{AB}  \s.
\end{equation} 
This state is distillable.

But what if we mix in more than two different EPR states?  Namely, instead of only considering 
hiding states (flags) correlated to odd parity Bell states (anti-key states) $|\psi_\pm\>={1\over \sqrt2}(|01\>\pm|10\>)$, we also add mix in
flags correlated to the even parity Bell states (key type states) $|\phi
_\pm \>={1\over \sqrt2}(|00\>\pm |11\>)$.
Consider:
\ben
&&\rho=p_{11}|\phi_+\>\<\phi_+|\ot \rho_{11} + 
p_{12}|\phi_-\>\<\phi_-|\ot \rho_{12} + \nonumber\\
&&+p_{21}|\psi_+\>\<\psi_+|\ot \rho_{21} +p_{22}|\psi_-\>\<\psi_-| \ot \rho_{22} 
\label{eq:nptboundstate}
\een
where 
\be
\rho_{ij}=\tau_i\ot \tau_j.
\ee
Let us take for example, all $p_{ij}=1/4$.  Then, after attempting to distinguish the hiding states, Alice and Bob will have
a state which is very close to the maximally mixed state (i.e. the state will be very close to a mixture of all four Bell states).  
The maximally mixed state is very far from being entangled, thus even if Alice and Bob's measurements on the hiding states are
able to bias the mixture away from the maximally mixed state, the state will still be separable.

Intuitively, it is thus clear why the state of equation (\ref{eq:nptboundstate}) will not be distillable.  Any protocol which attempts
to first distinguish which Bell state the parties have, will fail.  But is the state entangled?  Indeed it is, in fact it has
negative partial transpose.  To see this, we look at
the block-matrix form of the state 
\be
\rho={1\over 4} \left[
\begin{array}{cccc}
\tau_1 \ot (\tau_1+\tau_2) & 0 & 0& \tau_1 \ot (\tau_1-\tau_2)\\
0&\tau_2 \ot (\tau_1+\tau_2) & \tau_2 \ot (\tau_1-\tau_2)&0\\
0&\tau_2 \ot (\tau_1-\tau_2) & \tau_2 \ot (\tau_1+\tau_2)&0\\
\tau_1 \ot (\tau_1-\tau_2) & 0 & 0& \tau_1 \ot (\tau_1+\tau_2)\\
\end{array}
\right]
\ee
If the matrix were PPT we would have in particular 
\be
\tau_1^\Gamma \ot (\tau_1^\Gamma+\tau_2^\Gamma) \geq 
\tau_2^\Gamma \ot (\tau_1^\Gamma-\tau_2^\Gamma)
\label{eq:npt-ppt}
\ee
We will argue that it is not true.
Let us recall that 
\ben
&& \tau_1^\Gamma= \left( {P_+^\perp \over d^2-1}\right)^{\ot k} \nonumber \\
&&\tau_2^\Gamma= \left({P_+^\perp\over d^2+d}+ {(1+d) P_+\over d^2+d} \right)^{\ot k}
\equiv \left( {P_+^\perp \over d^2+d}\right)^{\ot k} + R
\nonumber \\
&&\tau_1^\Gamma - \tau_2^\Gamma= \left[(P_+^\perp)^{\ot k}\left( {1\over (d^2 -1)^k} - 
{1\over (d^2 + d)^k}\right)  - R\right]\nonumber \\
&&\tau_1^\Gamma + \tau_2^\Gamma= \left[(P_+^\perp)^{\ot k}\left( {1\over (d^2 -1)^k} + 
{1\over (d^2 + d)^k}\right)  + R\right]\nonumber \\
\een
where we use notation from sec. XI.
To see that (\ref{eq:npt-ppt}) is not satisfied  we consider the following projector
\be
Q=I-(P_+^\perp)^{\ot k}
\ee
i.e. $Q$ is projector onto support of positive operator $R$.
Since $\tr (Q\tau_1^\Gamma)=0$, we have  
\be
\tr[( Q\ot (P_+^\perp)^{\ot k} )(\tau_1^\Gamma \ot (\tau_1^\Gamma+\tau_2^\Gamma))] =0
\ee
Moreover we have 
\ben
&&\tr (Q\ot (P_+^\perp)^{\ot k}) [\tau_2^\Gamma \ot (\tau_1^\Gamma-\tau_2^\Gamma)] = \nonumber \\ 
&&=\tr R 
\left( {1\over (d^2 -1)^k} -  {1\over (d^2 + d)^k}\right)
\een
The above quantity is strictly greater than zero for $d\geq 2$. 
Thus inequality (\ref{eq:npt-ppt}) is violated on projector $Q\ot (P_+^\perp)^{\ot k}$.

Now, it may be that there is a protocol which succeeds in distilling from the state (\ref{eq:nptboundstate}) which does
not rely on first performing a measurement to distinguish the hiding states.  However, even taking many copies of the
state, produces a state of the form
\be
\rho=\sum_i |\psi_i\>\<\psi_i|\otimes\rho_i
\ee
with the $\rho_i$ being binary strings encoded in hiding states and $\psi_i$ being the basis of maximally entangled states.
Thus the form of the state is invariant under tensoring.  There is thus a very strong intuition that these states are NPT
bound entangled, and a very good understanding of why they might be so.  Effectively, the partial transpose does not feel very 
strongly the fact that the states $\rho$ are hiding states, but more strongly feels the fact that they are globally orthogonal.

\section{Controlled private quantum channels}
\label{sec:cpqc}

Here, we demonstrate a cryptographic application of bound entangled states which have key.  
A private quantum channel (PQC)\cite{boykin-qe,mosca-qe} allows for the sending of quantum states
such that an eavesdropper learns nothing about the sent states.
Here, we consider the cryptographic
primitive of having the ability to securely send quantum states (a PQC), but that this ability can be turned on and off by a 
controller.  Namely, we consider a three party scenario (Alice, Bob, and the (C)controller) and demand
\begin{itemize}
\item Alice and Bob have a private quantum channel, which they can use to send an unknown qubit from one to the other
in such a way that they can be sure that no eavesdropper (including the Controller), can gain information about the state being sent.
\item the Controller has the ability to determine whether or not Alice and Bob can send the qubit
\end{itemize}

We now show that this can be done using shared quantum states in such a way that the Controller only needs 
to send classical communication to one of the parties in order to activate the channel.  First, let us note that
the standard way of controlling the entanglement of two parties is via the GHZ state
\begin{equation}
|\psi\>_{ABC}=|000\>+|111\> \s .
\end{equation}
If the Controller, (Claire), measures in the basis $|0\>\pm |1\>$, then, depending on the outcome, Alice and Bob
will share either the Bell state $|\psi_+\>$ or $|\psi_-\>$.  If $C$ then tells them the result, they will have
one unit of entanglement (ebit) which they can then use to teleport quantum states. However, if the Controller wants
to give them the ability to send a single qubit securely, then the GHZ state cannot be used for this, because
the Controller can trick Alice and Bob into sending part of the quantum state to her.  She can claim that she
obtained measurement outcome $+$, when in reality she has not performed a measurement at all.  Then, when
Alice attempts to teleport a qubit to Bob, she is in fact teleporting to both Bob and the Controller.  The controller
can then perform a measurement on her qubit to obtain partial information about the sent qubit.
Note that here we are concerned with the ability to give single shot access to a quantum channel.  If the controller
gives Alice and Bob many ebits by performing measurements on many copies of a GHZ state, then Alice and Bob could always
perform purity testing to determine that the Controller is honest.

Let us know show that unlike the GHZ, the states of Eq. (\ref{he}) can be used in such a way that the Controller can give Alice and
Bob single shot access to a private quantum channel, in such a way that Alice and Bob are sure that the Controller cannot obtain
any information about the sent states even when the Controller cheats.  We will then show that we can do the same thing with fully
bound entangled states, so that Alice and Bob possess no distillable entanglement unless the Controller gives it to them.

First, we assume the shared state as a trusted resource.  I.e. a trusted party gives Alice, Bob and the Controller some
state which they use to implement the primitive.  This assumption can be removed in the limit of many copies, since if
Alice and Bob have many copies of the state, they can perform tomography to ensure that they indeed possess the correct state.
The state we initially use is the purification of Eq. (\ref{he})
\begin{equation}
\rho_{he} = {1\over 2} |\psi_+\>\<\psi_+|_{AB} \otimes \tau_1^{A'B'} + {1\over 2}|\psi_-\>\<\psi_-|_{AB} \otimes \tau_2^{A'B'} 
\end{equation}
Namely, 
\begin{equation}
|\psi\>_{ABC}=|00\>_{AB}\otimes|\phi_1\>_{A'B'C}+|11\>_{AB}\otimes|\phi_2\>_{A'B'C} 
\label{eq:cpqc}
\end{equation}
such that $\tr_C(|\phi_i\>\<\phi_i|)=\tau_i$. 

Thus, $\<\phi_1|\phi_2\>=0$ and since the $\tau_i$ are orthogonal, the Controller's states $\tr_{A'B'}(|\phi_i\>\<\phi_i|)=\sigma_C^i$
will be orthogonal.  The controller can thus give Alice and Bob one ebit by performing a measurement to distinguish the $\sigma^i_C$.
She then tells Alice and Bob the result.
Alice and Bob on the other hand, are guaranteed security by the fact that they either possess the state $|\psi_+\>$ or
$|\psi_-\>$.  I.e. it is an {\it incoherent} mixture of the two states, and they either have one of the states or the other,
they just don't know which one they have.  

The state of Eq. (\ref{he}) however, does have an arbitrarily small amount of distillable entanglement.  Thus, Alice and
Bob will have access to a private quantum channel in the case of having many copies of the state.   If we want
to give full control to Claire, we need to ensure that the state held by Alice and Bob in the absence of 
Claire's communication is non-distillable.  This can be achieved by using the bound entangled states of equation
(\ref{eq:rec-state}) which approximate a pbit.  It is not hard to verify, by explicitly writing the state in the
Bell basis on $AB$, that the state is arbitrarily close to a state of the same form as equation (\ref{eq:cpqc}),
and thus has the desired properties.

\section{Conclusion}
\label{sec:conc}
We have seen that one can recast obtaining a private key under LOPC in terms of distilling private states under LOCC.  
One finds a general class of states which are unconditionally secure.  This class includes bound entangled states from
which one cannot distill pure entanglement.  This then enables one to use tools developed in entanglement theory to
tackle privacy theory.  For example, the regularized relative entropy of entanglement was found to be an upper bound on the
rate of private key.

Many open questions remain. The most important problem in this context is whether all entangled states have non-zero distillable key or opposite - if there are bound entangled states which cannot be distilled into private states. 
 One can also ask about the private state cost
$K_{qc}$ of states $\rab$.  I.e. what is the dimension $d$ of the \main\ of the pdit that is required to 
create $\rab$ under LOCC?  It might even be that $K_{qc}=K_d$, which would enable entanglement theory to have basic
laws along the lines of \cite{thermo-ent2002}.

The question of reversibility of creating  states from private states touches another "qualitative" problem, namely how tight is the upper bound on distillable key which is the regularised relative entropy of entanglement.  

Exploring the wide class of private states especially in the context of the well established theory of distillation of entanglement appears to be
a necessary step in order to solve the above important problems.

\section{Appendix}
\subsection{Derivation of formula (\ref{eq:state_block_form}) of Sec. \ref{sec:eve-states}}
\label{subsec:App_Eves_states}

We show that a bipartite state of four subsystems $ABA'B'$ given as
\be
\rho_{ABA'B'}=\sum_{ij i'j'} |ij\>_{AB}\<i'j'| \ot A^{iji'j'}_{A'B'},
\ee
where $A_{iji'j'}$ are block matrices can be written as 
\be
\rho_{ABA'B'}=\sum_{iji'j'} \sqrt{p_{ij}p_{i'j'}}|ij\>_{AB}\<i'j'|\ot
[\ucal_{ij} \sqrt{\rho_E^{ij}}
\sqrt{\rho_E^{i'j'}}\ucal_{i'j'}^\dagger ]^T.
\ee

To see this, we first write down its total purification:
\be
\psi_{ABA'B'E}=\sum_{ij}\sqrt{p_{ij}}\,|ij\>_{AB} |\psi_{ij}\>_{A'B'E}.
\ee
The states $\psi^{ij}_{A'B'E}$ can be written as
\be
\psi^{ij}_{A'B'E}=\sum_{k=1}^{d_{A'B'}}\lambda^{ij}_k V_{ij} |k\>_{A'B'}\ot
U_{ij} \wcal |k\>_{A'B'}
\ee
Here $U_{ij}$ is unitary transformation acting on Eve's system,
$V_{ij}$ is unitary transformation acting on \side\ $A'B'$ and
$\wcal$ is some fixed embedding of $\hcal_{A'B'}$ into $\hcal_E$
(this is needed if Eve's systems are greater than the system $A'B'$):
\be
\wcal:\hcal_{A'B'}\to \hcal_E,\quad \wcal|k\>_{A'B'} =|k\>_E
\ee
where $|k\>_{A'B'}, k=1, \ldots, d_{A'B'}$ is a fixed basis in system $A'B'$,
while $|k\>_{E}, k=1, \ldots, d_{E}$ is a fixed basis in system $E$.
We will also need a dual operation, which is fixed projection of space
$\hcal_E$
into $\hcal_{A'B'}$:
\be
\wcal^\dagger:\hcal_{E}\to \hcal_{A'B'},
\ee
with
\ben
&&\wcal^\dagger|k\>_{E} =|k\>_{A'B'}\quad \mbox{for}\quad k=1,\ldots,
d_{A'B'} \\
&&\wcal^\dagger|k\>_{E} = 0 \quad \mbox{for} \quad k>d_{A'B'}
\een
One then finds that
\be
(A_{iji'j'})^T= V_{ij}^\dagger \wcal^\dagger U_{ij}^\dagger
\sqrt{\rho_E^{ij}}
\sqrt{\rho_E^{i'j'}} U_{i'j'}\wcal V_{i'j'}
\ee
where $T$ is matrix transposition. One can find,
the operator $\ucal_{ij}^\dagger\equiv U_{ij}\wcal V_{ij}$ maps the space
$\hcal_{A'B'}$
exactly onto a support of $\rho_E^{i'j'}$ in space $\hcal_E$, and the dual
operator
$\ucal_{ij}=V_{ij}^\dagger \wcal^\dagger U_{ij}^\dagger$
maps the support of $\rho_E^{ij}$ back to $\hcal_{A'B'}$. Finally, our state
is of the form
\be
\rho_{ABA'B'}=\sum_{iji'j'} \sqrt{p_{ij}p_{i'j'}}|ij\>_{AB}\<i'j'|\ot
[\ucal_{ij} \sqrt{\rho_E^{ij}}
\sqrt{\rho_E^{i'j'}}\ucal_{i'j'}^\dagger ]^T,
\ee
which we aimed to show.

\subsection{The proof of lemma \ref{lem:pptfam}, Section \ref{subsec:new-family}.}
\label{subsec:App_proof_lem:pptfam}
We prove now that the states from a family that we have introduced in eq. (\ref{row-key}), are indeed PPT for certain range of parameters, as it is stated in lemma \ref{lem:pptfam}.

{\proof 
The matrix of the state (\ref{row-key}) after partial transposition has a form
\be
\rho_{ABA'B'}^\Gamma=\left[\bea{cccc}
{p}({\tau_1+\tau_2\over 2})^{\Gamma} &0&0&0 \\
0& ({1\over 2} - p)\tau_2^{\Gamma}&{p}{({\tau_1-\tau_2\over 2}})^{\Gamma}&0 \\
0&p({{\tau_1-\tau_2}\over 2})^{\Gamma}&({1\over 2} -p)\tau_2^{\Gamma}& 0\\
0 &0&0& {p}({\tau_1+\tau_2\over 2})^{\Gamma}  \\
\eea
\right].
\label{aftgamma}
\ee
Since $\tau_1$ and $\tau_2$ are separable (and hence PPT), so is their mixture. Thus  extreme-diagonal
blocks of the above matrix are positive. It remains to check positivity 
of the middle block matrix. Since any block matrix of the form
\be
\left[\bea{cc}
A&B\\
B &A\\
\eea
\right],
\ee
is positive if there holds $A \geq |B|$ where $A$ and $B$ are 
arbitrary hermitian matrices, our question of positivity
of (\ref{aftgamma}) reads
\be
({1\over 2} - p)\tau_2^{\Gamma} \geq p|{({\tau_1-\tau_2\over 2}})^{\Gamma}|
\ee
Having  $\rho_s = {1\over d^2 +d}(I+V)$ and $\rho_a = {1\over d^2 - d} (I-V)$ 
where $V$
swaps d-dimensional spaces and applying $V^{\Gamma}= dP_+$ one easily gets that  
\ben
&& \tau_1^\Gamma= \left( {P_+^\perp \over d^2-1}\right)^{\ot k} \\
&&\tau_2^\Gamma= \left({P_+^\perp\over d^2+d}+ {(1+d) P_+\over d^2+d} \right)^{\ot k}
\label{tau2}
\een
where $P_+^\perp \equiv I-P_+$ is projector onto subspace orthogonal to the projector 
onto maximally entangled state $P_+ = |\psi_+\>\<\psi_+|$.

We check then  the  inequality 
\ben
&&({1\over 2} - p) \left({P_+^\perp\over d^2+d}+ {(1+d) P_+\over d^2+d} \right)^{\ot k}\geq \nonumber  \\
&&\geq {p\over 2}
\times \left|\left( {P_+^\perp \over d^2-1}\right)^{\ot k} - 
\left({P_+^\perp\over d^2+d}+ {(1+d) P_+\over d^2+d} \right)^{\ot k}\right|\nonumber \\
\een
To solve this inequality it is useful to represent the term on LHS as a sum:
\be
\left({P_+^\perp\over d^2+d}+ {(1+d) P_+\over d^2+d} \right)^{\ot k} 
 = \left({P_+^\perp\over d^2+d}\right)^{\ot k} + R
\ee
where operator $R$ is an unnormalised state which consists of all terms coming out 
of $k$-fold tensor product of 
$\left({P_+^\perp\over d^2+d}+ {(1+d) P_+\over d^2+d} \right)$ apart from 
the first term $\left({P_+^\perp\over d^2+d}\right)^{\ot k}$.
It is good to note that $R$ has support on subspace 
orthogonal to $(P_+^\perp)^{\ot k}$. This fact allows 
to omit the modulus and to get 
\ben
&&({1\over 2} - p)\left[ \left({P_+^\perp\over d^2+d}\right)^{\ot k} + R\right]\geq    \nonumber \\
&&\geq {p\over 2}\left[(P_+^\perp)^{\ot k}\left( {1\over (d^2 -1)^k} - 
{1\over (d^2 + d)^k}\right)  + R\right]
\een 
Since $R$ and $(P_+^\perp)^{\ot k}$ are orthogonal, this inequality is 
equivalent to the following two inequalities 
\ben
&&({1\over 2} - {3\over 2}p) R \geq 0 \\
&& ({1\over 2} - p) \left({P_+^\perp\over d^2+d}\right)^{\ot k} \geq 
  {p\over 2}(P_+^\perp)^{\ot k}\times \nonumber \\ 
&& \times \left( {1\over (d^2 -1)^k} - {1\over (d^2 + d)^k}\right)
\een
To save first inequality one needs $p \leq {1\over 3}$. Preserving the second one
requires
\be
{1- p\over p} \geq \left({d\over d-1}\right)^k
\ee
This however is fulfilled for any $p \in (0,{1\over 3}]$ if  $d$ is taken properly large for some fixed k. Indeed, the
$k$-th root of $1 -p \over p$ (which converges to $1$ with k) can be greater than $d\over d-1$ (which converges
to $1$ with d) for some large d. 
}\blacksquare

\subsection{Comparison of two criteria for secure key}
\label{sec:securitycrit}

In this section, we shall compare the joint cryptographic criterion,
i.e. the requirement of (\ref{eq:joint}):
\be
||\rho^{ccq}_{real} - \rho^{ccq}_{ideal}||\leq \ep
\label{eq:joint2}
\ee
which includes both uniformity and security in one formula 
with the double condition where uniformity and security are treated separately, namely:
\ben
\chi(\{p_i,\rho^E_{ij}\}) \leq \ep \label{eq:doub-cond}\\
||\rho_{AB} - \rho^{AB}_{ideal}|| \leq \ep \nonumber
\een
The connection between these two criteria for quantum cryptographical security of the state
is given in the theorem below. 

{\theorem  For any ccq state 
$\rho_{ABE} = \sum_{ij=0}^{d-1} p_{ij} |ij\>\<ij| \otimes \rho_{ij}^E$ and
$\rho_{ideal} = \sum_{i=0}^{d-1}{1\over d}|ii\>\<ii|\otimes \rho_E$ where $\rho_E = \sum_{ij} p_{ij}\rho_{ij}^E$,
the following implications holds:
\be \left. \bea{l}
|| \rho_{AB} - \rho^{AB}_{ideal}|| \leq \ep \\
\chi(\rho_{ABE}) \leq \ep
\eea
\right \}\Rightarrow ||\rho_{ABE} - \rho_{ideal}|| \leq \ep + \sqrt{\ep}
\label{firstpartlem}
\ee

\be
  ||\rho_{ABE} - \rho_{ideal}|| \leq \ep \Rightarrow\left \{ \bea{l}
 \chi(\rho_{ABE}) \leq 4\ep \log d + h(\ep)  \\
|| \rho_{AB} - \rho^{AB}_{ideal}|| \leq \ep.
\eea
\right.
\label{secpartlem}
\ee
where $\rho_{AB}=\tr_E \rho_{ABE}$, \, $\rho^{AB}_{ideal}=\tr_E \rho_{ideal}$.
\label{equivlemma}
}


{\rem 
We see that the result (\ref{secpartlem}) is not fully satisfactory due to the term $\log d$.
However, one cannot get a better result.  Indeed it is easy to construct a state,
for which the Holevo function is of $\ep \log d$ order, though the state $\rho_{ABE}$ is $\ep$ close
to some $\rho_{ideal}$ state. As an example may serve an appropriate extension of the isotropic state, measured
in computational basis:
\ben
&&\rho_{ABE} = (1-\ep)(\sum_{i=0}^{d-1}{1\over d}|ii\>\<ii|_{AB})\ot (|00\>\<00|)_E  +  \nonumber \\ 
&&+ \ep \sum_{i \neq j}{1\over d^2 - d}|ij\>\<ij|_{AB}\ot (|ij\>\<ij| )_E 
\een
where $\sigma$ is maximally mixed state. If we consider now 
$\rho_{ideal} = (\sum_{i=0}^{d-1}{1\over d}|ii\>\<ii|_{AB})\ot ( |00\>\<00|)_E$, 
it is easy to see that $||\rho_{ABE} - \rho_{ideal}|| = 2\ep$.  However the value of the Holevo function 
equals $h(\ep) + \ep \log(d^2 - d)$. 
}

{\rem
The main difficulty in the proof of the above theorem is to get the term $\log d$ ($d\times d$ is size of $AB$ system) rather than $\log d_{ABA'B'}$. 
The latter one would be obtained directly from Fannes type continuities. 
However to get $\log d$ we have to apply tricks based on twisting. 
It is quite convenient not to have Eve's dimension in equivalence formula. This is because Eve's dimension 
depends on the protocol that lead to the key (more specifically, it depends on the amount of communication). In contrast, 
dimension of Alice and Bob system is only the number of bits of obtained key.
Thus, our equivalence is independent of the protocol. 
}

{\proof 
For the first part of the theorem \ref{equivlemma} we assume that 
 $\chi(\{p_{ij},\rho_{ij}^E\})\leq \ep$ which
by proposition \ref{secbounds} in Sec. \ref{subsec:app-secur} of Appendix
means that we have:
\be
||\sum_{i,j=0}^{d-1} p_{ij} |ij\>\<ij|_{AB}\otimes\rho_{ij}^E - 
\sum_{i,j=0}^{d-1} p_{ij} |ij\>\<ij|_{AB}\otimes\rho_E|| \leq \sqrt{\ep},
\label{proofOfEquivalBeg}
\ee
with $\rho_E = \sum_{i,j=0}^{d-1}p_{ij} \rho_{ij}^E$.
Moreover by second assumption that 
\be
||\sum_{i,j=0}^{d-1} p_{ij} |ij\>\<ij| - \sum_{i}^{d-1} {1\over d} |ii\>\<ii|) || \leq \ep
\ee
one gets 
\ben
||\sum_{i,j=0}^{d-1} p_{ij} |ij\>\<ij|_{AB}\otimes\rho_E - 
\sum_{i=0}^{d-1} {1\over d} |ii\>\<ii|_{AB}\otimes\rho_E|| \leq \ep.
\label{eq:lipa}
\een
Using triangle inequality, and Eqs.  (\ref{proofOfEquivalBeg}) and (\ref{eq:lipa})
one obtains the 
\be
||\rho_{ABE} - \rho_{ideal} ||\leq \ep + \sqrt{\ep}.
\label{ABE-ideal}
\ee

The proof of the second part of the theorem \ref{equivlemma} is a bit more involved. Of course,  it is immediate that due to monotonicity
of trace norm under partial trace, from $||\rho_{ABE}- \rho_{ideal}||\leq \ep$ it follows $||\rho_{AB}- \rho_{ideal}^{AB}||\leq \ep$.
The non-obvious task is to bound also $\chi$. 
So, we assume that
\be
||\rho_{ABE}-\rho_{ideal}||\leq \ep,
\label{someineq}
\ee
By equality of norm and fidelity condition (\ref{fidnor}), there holds 
\be
F(\rho_{ABE},\rho_{ideal}) \geq 1 -{1\over 2}\ep
\ee
By definition of fidelity, there are pure states $\psi$ and $\phi$ (purifications of 
$\rho_{ABE}$ and $\rho_{ideal}$ respectively), such that $F(\psi,\phi) = F(\rho_{ABE},\rho_{ideal})$.
Without loss of generality we can consider the system which purifies both states  to be bipartite. We will call
it $A'B'$. Now let us perform twisting operation on the $ABA'B'$ parts of the pure states $\psi$ and $\phi$,
 which in the case of state $\rho_{ideal}$ transforms $AB$ subsystem of into maximally entangled state - $P_d^+$, 
(we can choose such twisting because by the theorem \ref{th:pdit} purification of an ideal state is some pdit state). 
I.e. after such twisting, pdit will become a basic pdit (\ref{eq:ibasicpdit}) which is product with $A'B'$ subsystem.
Since unitary transformation and tracing out can only increase fidelity,  then 
applying again (\ref{fidnor}) we have that  subsystem $AB$ of $\rho_{ABE}$  is close to a singlet state in norm:
\be
||\rho_{AB}-P_d^+||\leq 2\sqrt{\ep}.
\label{ineqii}
\ee
Using Fannes inequality (see eq. \ref{eq:Fannes_ineq} in Sec. \ref{subsec:notation}) we get 
\be
S(\rho_{AB})  \leq 4\sqrt{\ep}\log d  + h(2\sqrt{\ep}).
\ee 
Since the total state of systems $ABA'B'E$ is pure, 
we get that $S(\rho_{A'B'E}) = S(\rho_{AB})$ hence 
\be
S(\rho_{A'B'E})\leq 4\sqrt{\ep}\log d  + h(2\sqrt{\ep}).
\label{ineq1}
\ee
Now, note  that the state of the system  $A'B'E$ has the  form
\be
\rho_{A'B'E} = \sum_{i,j=0}^{d-1}p_{ij}\rho_{ij}^{A'B'E}, 
\ee
where the state $\rho_{ij}^{A'B'E}$ denotes state of $A'B'E$ system after twisting and  
given that $AB$ subsystem is in state $|ij\>\<ij|$ (i.e. if after twisting one measure the system $AB$ in basis $|ij\>$ 
the system  $A'B'E$ would collapse to $\rho_{ij}^{A'B'E}$).  By definition of Holevo function there holds:
\be
\chi(\{p_{ij},\rho_{ij}^{A'B'E}\}) \leq S(\rho_{A'B'E}).
\label{ineq2}
\ee

The question is how the Holevo function of the $\{p_{ij},\rho_{ij}^{A'B'E}\}$
ensemble is related to Holevo function of  $\{p_{ij},\rho_{ij}^{E}\}$ which we would like to bound from above.
It is crucial, that by theorem \ref{th:ccq} twisting operation does not affect 
the ccq state which comes out of the measurement of $AB$ in control basis of twisting. In other words, the ensemble $\{p_i,\rho^E_{ij}\}$ does not 
change under twisting, so that $\rho^E_{ij}=\tr_{A'B'}\rho^{A'B'E}_{ij}$.
It is easy now to compare the functions $\chi(\{p_{ij},\rho_{ij}^{E}\})$ and $\chi(\{p_{ij},\rho_{ij}^{A'B'E}\})$:
\be
\chi(\{p_{ij},\rho_{ij}^{E}\})  \leq \chi(\{p_{ij},\rho_{ij}^{A'B'E}\}) 
\label{ineq3}
\ee
This is due to the fact, that each state $\rho_{ij}^{E}$ can be obtained from $\rho_{ij}^{A'B'E}$ by 
tracing out $A'B'$ subsystem. However tracing out can only decrease Holevo function, because
this function is equal to the average relative entropy:
\be
\chi(\{p_k,\rho_k\})=\sum_k p_k S(\rho_k |  \sum_k p_k \rho_k).
\ee
Summing up the chain of inequalities (\ref{ineq1}), (\ref{ineq2}) and (\ref{ineq3}) one gets
\be
\chi(\{p_{ij},\rho_{ij}^{E}\})  \leq 4\sqrt{\ep}\log d  + h(2\sqrt{\ep}).
\label{woowineq}
\ee
which is a desired security condition - bound on the Holevo function of the ansamble $\{p_{ij},\rho_{ij}^{E}\}$. }
\blacksquare

\subsection{Useful inequalities relating security conditions}
\label{subsec:app-secur}
In this section we collect  relations between 
different security conditions for \ccq states. Some of 
these relations have been studied in \cite{Ben-OrHLMO05}. 
Since we will not deal with uniformity, but solely with security, 
it is convenient to use single index $k$ in place of $ij$. 
We thus consider \ccq state  (which could be actually called cq state)
\be
\rho = \sum_{k=0}^{d^2-1}p_k |k\>\<k|\otimes\rho_k 
\label{real}
\ee

Basing on this fact, we can state another lemma establishing some equivalences:
{\lemma
For any state (\ref{real}) and any positive real $\ep\leq {1\over 2}$, 
the following implications hold
\begin{enumerate}
\item 
\ben
\label{normdist} 
&&||\sum_k p_k |k\>\<k|\otimes\rho_k - (\sum_j p_j |j\>\<j|)\otimes\rho||\leq \ep  
\\
&&\Rightarrow 
\sum_k p_k F(\rho_k,\rho) \geq 1 - {1\over 2}\ep \nonumber 
\een
\item 
\be
\label{fiddist} 
\sum_k p_k F(\rho_k,\rho) \geq 1-\ep 
\Rightarrow 
\sum_k p_k ||\rho_k - \rho||\leq 8\ep
\nonumber 
\ee
\item 
\ben 
\label{avnormdist} 
&&\sum_k p_k ||\rho_k - \rho||\leq \ep  \Rightarrow \\
&&\Rightarrow 
||\sum_k p_k |k\>\<k|\otimes\rho_k - (\sum_j p_j |j\>\<j|)\otimes\rho|| \leq \ep.
\nonumber
\een
\label{lemprzedost}
\end{enumerate}
Here  $\rho=\sum_k p_k \rho_k$. 
\label{lem:sec-fid-norm}
}

{\proof 
The first thesis follows from the mentioned equivalence of norm and fidelity
and definition of fidelity. Namely one can make use of 
lemma \ref{lemfidnor}, so that if (\ref{normdist}) holds, the
fidelity $F\big(\sum_k p_k |k\>\<k|\otimes\rho_k,(\sum_j p_j |j\>\<j|)\otimes\rho\big)$ 
is no less than $1 -{1\over 2}\ep$. However it  is equal to 
average fidelity $\sum_k p_k F(\rho_k,\rho)$. 
Indeed, 
\be
F(\sum_k p_k |k\>\<k|\otimes\rho_k,(\sum_j p_j |j\>\<j|)\otimes\rho)=  
\tr \biggl( \bigl( \sum_k p_k |k\>\<k|\otimes \rho_k \bigr)^{1\over 2}
\sum_j p_j |j\>\<j|\otimes\rho \biggr. 
\biggl.\bigl(\sum_l p_l |l\>\<l|\otimes\rho_l \bigr)^{1\over 2} \biggr)^{1\over 2}
\ee
Now by orthogonality of vectors $|k\>$ one 
has 
\be
\sqrt{\sum_k p_k |k\>\<k|\otimes\rho_k}=\sum_k 
\sqrt{p_k} |k\>\<k|\otimes\sqrt{\rho_k}.
\ee
Multiplying now the $(\sum_j p_j |j\>\<j|)\otimes\rho$ matrix
by the above from left-hand-side and right-hand-side one gets
\be
\sum_k p_k^2 |k\>\<k|\otimes 
\sqrt{\rho_k}\rho \sqrt{\rho_k}.
\ee 
This immediately gives the above formula equal to
\be
\sum_k p_k \tr\sqrt{\sqrt{\rho_k}\rho\sqrt{\rho_k}}
\ee
which is just average fidelity from (\ref{fiddist}).

The second thesis of this lemma ( Eq. (\ref{fiddist})) is again a consequence 
of (\ref{fidnor}). If applied to each pair $\rho_k$, $\rho$, 
and averaged over probabilities of $p_k$ gives that 
\be
\sum_{k} p_k \sqrt{1- {1\over 4}||\rho - \rho_k||^2} \geq 1-\ep
\ee
which is equivalent to
\be
\sum_{k} p_k \left(1 - \sqrt{1- ||\rho - \rho_k||^2/4} \right)\leq \ep.
\ee
Now by the fact that
\be
1 - \sqrt{1- {1\over 4}||\rho - \rho_k||^2}  
\ee
is a convex function of $||\rho - \rho_k||$ on interval $(0,2)$ we get
\be
1 - \sqrt{1- {1\over 4}{[\sum_{k} p_k||\rho - \rho_k||]}^2} \leq \ep
\ee 
This however reads for $0 <\ep < 1 $ 
\be
\sum_{k} p_k||\rho - \rho_k||\leq 8\ep.
\ee
Since $||\rho - \rho_k||\leq 2$ one has, that for $\ep \geq 1$ the above inequality
is also valid, which completes the proof of the second thesis of lemma \ref{lem:sec-fid-norm}.

The last implication 
(Eq. (\ref{avnormdist})) is a consequence of triangle inequality,
which completes the lemma.}\blacksquare

Let us notice, that this lemma establishes a kind of equivalance of security 
conditions, namely:
\ben
||\sum_k p_k |k\>\<k|\otimes\rho_k - (\sum_j p_j |j\>\<j|)\otimes\rho||\leq \ep \\
\Rightarrow \nonumber
 \sum_{k} p_k||\rho - \rho_k||\leq 4\ep \\
 \Rightarrow \nonumber
||\sum_k p_k |k\>\<k|\otimes\rho_k - (\sum_j p_j |j\>\<j|)\otimes\rho||\leq 4\ep
\een

We can show now links between the above conditions on ccq state and Holevo 
function $\chi$ of this state, i.e. of an ansamble $\{p_k, \rho_k\}$ which we shall
write $\chi(\rho_{ccq})$.

{\lemma For any ccq state $\rho_{ccq}$ (\ref{real}) there holds:
\ben
&&\chi(\rho_{ccq})\leq \ep \Rightarrow \sum_k p_k ||\rho_k - \rho||\leq \sqrt{2\ep}  \\ \nonumber
&&\sum_k p_k ||\rho_k - \rho||\leq \ep \Rightarrow  
\chi(\rho_{ccq})\leq 3\ep \log d + \max (h(\ep),2\ep) \nonumber 
\een
where $\sum_k p_k\rho_k=\rho$, which acts on Hilbert space  $\hcal = {\cal C}^d$, 
and $h(\ep)=-\ep \log \ep - (1-\ep) \log (1-\ep) $ is binary entropy.
\label{lemost}
}

{\proof 
For the first statement of this lemma, 
let us notice that $\chi(\rho_{ccq})=S(\rho) - \sum_k p_k S(\rho_k)$ 
is just equal to average relative entropy distance 
$\sum_k p_k S(\rho_k | \rho)$. Thus, by assumption we have 
\be
 \chi(\rho_{ccq}) =\sum_k p_k S(\rho_k | \rho) \leq \ep.
\label{lem:ass}
\ee
Now we can make use of the inequality \cite{OhyaPetz}:
\be
{1 \over 2} ||\rho - \rho_k||^2 \leq S(\rho_k|\rho)
\label{normerdist2}
\ee
which after averaging over probabilities and by concavity of square root 
gives
\be
\sum_k p_k ||\rho - \rho_k|| \leq \sqrt{2 \sum_k p_k S(\rho_k |\rho)}.
\ee
Applying now bound (\ref{lem:ass}) we obtain

\be
\sum_k p_k ||\rho - \rho_k|| \leq \sqrt{2\ep}
\ee
which completes first thesis of this lemma.

To prove the second statement of the lemma we use the Fannes inequality (see eq. \ref{eq:Fannes_ineq} in Sec. \ref{subsec:notation}). 
Namely, for $||\rho - \rho_k||\leq 1$ there holds:
\be
|S(\rho)-S(\rho_k)|\leq 2||\rho-\rho_k||\log d + h(||\rho-\rho_k||).
\ee
Let $G =\{k : ||\rho -\rho_k||\leq 1\}$ and $B =\{k : ||\rho - \rho_k|| >1\}$, and denote 
$\sum_{k\in G} p_k[S(\rho)-S(\rho_k)] \equiv \chi_G(\rho_{ccq})$, and $\chi_B(\rho_{ccq})$ analogously. We then have,
that 
\be
\chi(\rho_{ccq})=\chi_G(\rho_{ccq})+\chi_B(\rho_{ccq}).
\ee

We will give now the bounds for $\chi_G(\rho_{ccq})$ and $\chi_B(\rho_{ccq})$ respectively.
The first quantity is directly bounded by the analogous sum over LHS of the Fannes inequality:
\be
\chi_G(\rho_{ccq}) \leq \sum_{k\in G} p_k \bigl( 2||\rho - \rho_k||\log d + h(||\rho -\rho_k||)\bigr).
\ee
>From assumption 
\be
\sum_k p_k ||\rho-\rho_k||\leq \ep
\label{eq:curr_ass}
\ee
it follows that $\sum_{k\in G} p_k||\rho-\rho_k|| \leq \ep$. Using this, and adding non-negative terms $\sum_{k\in B} p_k h(||\rho -\rho_k||)$, we get:
\be
\chi_G(\rho_{ccq}) \leq 2\ep\log d + \sum_k p_k h(||\rho -\rho_k||).
\ee
Now by concavity of binary entropy one gets
\be
\chi_G(\rho_{ccq}) \leq 2\ep\log d + h( \sum_k p_k||\rho-\rho_k||),
\ee
Were the entropy increasing on $[0,\infty ]$ interval, one could use directly 
the assumption that $\sum_k p_k||\rho-\rho_k|| \leq \ep$, and bound $h( \sum_k p_k||\rho-\rho_k||)$  by $ h(\ep)$. Since it is the case only for $\ep \in [0,{1\over 2}]$, 
we have to end up with more ugly, but nonetheless useful expression. 
Namely on the interval $({1\over 2}, \infty]$ where the entropy becomes decreasing,
it is bounded by 1, and hence not greater than $2\ep$ for $\ep \in ({1\over 2}, \infty]$.
Thus finally one gets 
\be
\chi_G(\rho_{ccq})\leq 2\ep \log d + \max (h(\ep),2\ep),
\label{eq:G-entr-bound}
\ee

We turn now to give the bound for $\chi_B(\rho_{ccq})$. For the latter we use the fact, that the Holevo quantity is bounded from above by $\log d$, which gives:
\be
\chi_B(\rho_{ccq}) \leq \sum_{k\in B} p_k \log d.
\ee
To bound the last inequality we observe, that by definition of the set $B$ we have $\sum_{k\in B} p_k||\rho -\rho_k|| \geq \sum_{k\in B} p_k$. Then, again by assumption (\ref{eq:curr_ass}), we have 
\be
\chi_B(\rho_{ccq}) \leq \ep \log d.
\ee
Collecting inequality (\ref{eq:G-entr-bound}) and the above one, we arrive at the formula
\be
\chi(\rho_{ccq})\leq 3\ep \log d + \max (h(\ep),2\ep),
\label{eq:entr-bound}
\ee
which ends the proof of the lemma.
} \blacksquare

The lemmas above allow to prove the following proposition
{\prop For any ccq state (\ref{real}) 
the following holds:
\ben
\chi(\{p_k,\rho_{k}\})\leq \ep
\Rightarrow\nonumber \\
||\sum_k p_k |k\>\<k|\otimes\rho_k - (\sum_j p_j |j\>\<j|)\otimes\rho||\leq \sqrt{2\ep}
\een
where $\rho = \sum_k p_k \rho_k$. 
\label{secbounds}
}

{\proof 
Assuming that Holevo  function is smaller than $\ep$, we get by lemma \ref{lemost} that 
$\sum_k p_k ||\rho_k - \rho|| \leq \sqrt{2\ep}$. This however implies
by lemma (\ref{lem:sec-fid-norm}) that 
$||\sum_k p_k |k\>\<k|\otimes \rho_k - (\sum_j p_j |j\>\<j|)\otimes \rho ||$ is also
not greater than $\sqrt{2\ep}$, which completes proof of  the proposition.}  \blacksquare


\subsection{Properties of \pbits}
\label{subsec:app_prop_of_pbits}
We shall give here detailed proof of the lemma \ref{lem:negofpbit}, Section \ref{ssec: pbtexpls}.

{\proof  Log-negativity \cite{Vidal-Werner} (cf. \cite{ZyczkowskiHSP-vol}) is 
defined as $E_N(\rho)=\log (||\rho^{\Gamma}||)$. 
It is easy to see, that after partial transposition on $BB'$ subsystem,
the \pbit\ $\gamma$ in $X$-form changes into
\be
\gamma_{ABA'B'}^{\Gamma}={1\over 2}\left[\bea{cccc}
\sqrt{XX^{\dagger}}^{\Gamma} &0&0&0 \\  
0& 0&X^{\Gamma}&0 \\
0&(X^{\dagger})^{\Gamma}&0& 0\\
0&0&0&\sqrt{X^{\dagger}X}^{\Gamma}\\
\eea
\right].
\ee
We have 
\be
||\gamma^\Gamma||= {1\over 2} (||[\sqrt{XX^\dagger}]^{\Gamma}||+||[\sqrt{X^\dagger X}]^{\Gamma}||
+||A||),
\ee
where
\be
A =\left[\bea{cc}
0&(X)^{\Gamma} \\  
(X^{\dagger})^{\Gamma}&0 \\
\eea
\right].
\ee
By assumption, the operators $[XX^\dagger]^{\Gamma}$ and $[X^\dagger X]^{\Gamma}$ are positive, so that 
\be
||[\sqrt{XX^\dagger}]^{\Gamma}||+||[\sqrt{X^\dagger X}]^{\Gamma}||=\tr (\sqrt{XX^\dagger}+\sqrt{X^\dagger X})^{\Gamma}
=2\tr \gamma^\Gamma=2.
\ee
The last equality comes from the fact that $\Gamma$ preserves trace.
To evaluate norm of $A$, we note that  due to unitary invariance of 
trace norm we have $||A||=||\sigma_x\ot I_{A'B'} A||$.
Consequently 
\be
||A||= ||X^\Gamma||+||(X^\dagger)^\Gamma||=2 ||X^\Gamma||.
\ee
The last equality follows form the fact that $\Gamma$ commutes with Hermitian 
conjugation, and trace norm is invariant under Hermitian conjugation $||X||=||X^\dagger||$. 
Thus we get 
\be
E_N(\gamma)=\log(1+||X^\Gamma||),
\ee
which proves the lemma.\blacksquare
}
\subsection{Relative entropy of entanglement and \pdits}
\label{app:relentpdit}

We give now the proof of the theorem \ref{th:regrelentpbit},
Section \ref{app:relentpdit}.

{\proof  If one consider the thesis of theorem \ref{th:erpbit} for the state $\rho= \gamma_{ABA'B'}^{\ot n}$, it follows that
\be
E_r(\rho)\leq  \log d^n + {1\over d^n}\sum_{k = 0}^{d^n-1} E_r(\sigma_k),
\ee
with $k$ being the multiindex $k = (i_1,...,i_n)$ with $i_l \in \{0,...,d-1\}$ for $l \in \{1,...,n\}$ and
$\sigma_k = \rho_{i_1}\ot ...\ot \rho_{i_n}$.
Dividing both sides by $n$ we obtain
\be
{1\over n}E_r(\gamma_{ABA'B'}^{\ot n})\leq  \log d + {1\over n d^n}\sum_{k = 0}^{d^n-1} E_r(\sigma_k),
\ee
The left-hand-side of this inequality approaches $E_r^{\infty}(\gamma)$ with $n$. What has to be shown is 
that 
\be
\lim_{n\rightarrow \infty}{1\over n d^n}\sum_{k = 0}^{d^n-1} E_r(\sigma_k) \leq
\sum_{i=0}^{d-1} {1\over d} E_r^{\infty}(\rho_i),
\label{eq:stronglow}
\ee
with $\rho_i$ denoting the conditional states on $A'B'$ subsystem. 

Let us first observe that 
$E_r(\sigma_k) = E_r(\sigma_{k'})$ for any $k$ and $k'$ which are of the same type, i.e. which has the same numbers of occurrence of symbols from set $\{0,...,d-1\}$. This is because $\sigma_k$ and $\sigma_k'$ differ by local reversible transformation which does not change the entanglement. 
Moreover, as we will see, one can consider only those $\sigma_k$ for which $k$ is $\delta$-strongly typical i.e. such, that for some fixed $\delta > 0$ there holds \cite{CoverThomas}:
\be
\forall_{a\in \{0,...,d-1\}} \quad | {a(k)\over n} - {1\over d}| < \delta,
\label{eq:signals}
\ee
where $a(k)$ denotes frequency of symbol $a$ in sequence $k$. The set of such $k$ of length $n$ we will denote as $ST_{\delta}^{n}$. It is known, that the strongly typical set carries almost whole probability mass  for large $n$, that is for any $\delta > 0$ and any $\ep>0$ there exists $n_0$ such that for all $n\geq n_0$:
\be
P^{\ot n}(ST^n_{\delta}) > 1-\ep.
\ee
Now, since we deal here with homogeneous distribution, we can say, that the probability of the set of events is directly related to the power of this set. Namely we have:
\be
|ST^n_{\delta}|/d^n > 1-\ep,
\ee
which gives $|ST^n_{\delta}| > (1-\ep)d^n$.

We can rewrite now the term of the LHS of (\ref{eq:stronglow}) as follows:
\be
{1\over n d^n}\sum_{k = 0}^{d^n-1} E_r(\sigma_k)=
 {1\over n d^n}\bigl(\sum_{k \in ST_{\delta}^n} E_r(\sigma_k)  +
\sum_{k \not \in ST_{\delta}^n} E_r(\sigma_k)\bigr).
\ee
We can get rid of the second term of the RHS of the above equality, because we can bound from above 
for each $k$ the term $E_r(\sigma_k)$ by $n\log d$, which gives:
\be
{1\over n d^n}\sum_{k = 0}^{d^n-1} E_r(\sigma_k)\leq 
 {1\over n d^n}\bigl(\sum_{k \in ST_{\delta^n}} E_r(\sigma_k)  +
n \log d \sum_{k\not\in ST_{\delta}^n} \bigr) \leq 
{1\over n d^n}\sum_{k \in ST_{\delta}^n} E_r(\sigma_k)  +
\ep \log d
\ee
Now for each $k \in ST_{\delta}^n$ we have
\be
E_r(\sigma_k) = E_r(\rho_0^{\ot \tilde{m}_0}\ot \rho_1^{\ot {\tilde m}_1} \ot ... \ot \rho_{d-1}^{\ot {\tilde m}_{d-1}}) 
\ee
with ${\tilde m}_{i} = i(k)$ with $i$ in place of $a$ in (\ref{eq:signals}). 
By subadditivity of $E_r$ one has
\be
E_r(\sigma_k) \leq \sum_{l=0}^{d-1} E_r(\rho_l^{\ot {\tilde m}_l}).
\ee 
Note, that $\rho_l$ stands here for the state on shield part of {\it one} copy of $\gamma_{ABA'B'}$.
Applying this inequality for each $k$ in $ST_{\delta}^n$ and taking maximum of
LHS of the above inequality over $k$, we have a bound:
\be
{1\over n d^n}\sum_{k=0}^{d^n-1} E_r(\sigma_k) \leq 
  \sum_{l=0}^{d-1} {1\over n}E_r(\rho_l^{\ot m_l}) + \ep\log d,
\ee
where $m_l$ are the coefficients of the decomposition of some $\sigma_k$ into $\rho^{\ot m_l}$, which yields the maximal value of $E_r(\sigma_k)$ over all strongly typical $k$.
Now, we can rewrite the RHS of the above inequality as:
\be
\sum_{l=0}^{d-1} {m_l\over n}{1\over m_l}E_r(\rho_l^{\ot m_l}) + \ep\log d \leq 
({1\over d} + \delta)\sum_{l=0}^{d-1} {1\over m_l}E_r(\rho_l^{\ot m_l}) + \ep\log d 
\ee
where the last inequality holds for sufficiently high $n$ by assumption of strong typicality.
Thus, for every $\delta$ and $\ep$ and sufficiently large $n$ there holds:
\be
{1\over n d^n}\sum_{k = 0}^{d^n-1} E_r(\sigma_k) \leq 
({1\over d} +\delta)\sum_{l=0}^{d-1} {1\over m_l}E_r(\rho_l^{\ot m_l}) +\ep \log d,
\ee
One then sees, that the RHS approaches 
\be
({1\over d}+\delta)\sum_{l=0}^{d-1} E_r^{\infty}(\rho_l) + \ep\log d
\ee 
in limit of large $n$. Indeed, since for every $l$ the limit $E_r^{\infty}(\rho_l)$ exists, the subsequence ${1\over m_l}E_r(\rho_l^{\ot m_l})$ approaches this limit. Taking now infimum over $\delta$ and $\ep$ we prove the inequality (\ref{eq:stronglow}). This proves the theorem \ref{th:regrelentpbit}. \blacksquare
}

\subsection{Approximate pbits}
\label{app:approx-pbits}

We give here the proof of lemma \ref{lem:2qcoh}

{\proof
Assume first, that $\tr\rho_{AB}P_+> 1-\ep$. Since the elements $a_{ijkl}$ are real, by hermicity of the state
we have 
\be
\tr \rho_{AB}P_+ = {1\over 2}(a_{0000}+a_{1111} + 2Re(a_{0011})) 
\label{eq:trace}
\ee
This is however less than or equal to ${1\over 2}(1 +2a_{0011}) $, which is in turn
greater than $1-\ep$, and the assertion follows.
For the second part of the lemma, assume that $a_{0011}>{1\over 2}-\ep$.  We then have 

 $$\tr\rho_{AB}P_+ >{1\over 2}(a_{0000}+a_{1111} + 1 -2\ep). $$

We now bound the sum of $a_{0000}$ and $a_{1111}$. By positivity of the state, we have that 
$\sqrt{a_{0000}a_{1111}}>|a_{0011}|$. Now, by arithmetic-geometric mean inequality, we have that
$a_{0000}+a_{1111} \geq 2\sqrt{a_{0000}a_{1111}}$ which gives the proof. \blacksquare
 }

\subsection{Relative entropy bound}
\label{Appendix:proof_farfromsinglet}
{\proof   (of Lemma \ref{farfromsinglet}, Section \ref{sec:relent})
Let us first show, that 
\be
\tr P^+_d \sigma_{AB}\leq {1\over d}
\label{pplusfar}
\ee
for any $\sigma_{AB}\in T$. 
We first show this for $\sigma_{AB}$ "derived" from some 
pure product states $|\psi\>\<\psi|$: 
\be
\sigma_{AB}=\tr_{A'B'}U^{\dagger}|\psi\>\<\psi| U.
\ee
Because $\psi$  is product, it can be written as 
\beq 
\psi = (\sum a_i |i_A\> |\psi_i\>)
\otimes
(\sum b_i|i_B\> |\phi_i\>)
\eeq
with $a_i, b_i$ normalized and $|i_A\>,|i_B\>,|\psi_i\>,|\phi_i\>$ on
subsystem $A,B,A',B'$ respectively.

Now  the condition that the reduced $AB$ state 
has overlap with $P^+_d$ no greater than $1/d$ is 
\beq
\sum_{ij} a_i b_i a^*_j b^*_j \< x_i|x_j\> \leq 1
\eeq
where $x_k$ are arbitrary vectors of norm one arising from
the action of $U$ on $\psi_i$ and $\phi_i$. 
Since the $x_k$ are arbitrary they can incorporate the phases
of $a_i, b_i $
so that we require now
$\sum_{ij} \sqrt{p_iq_i p_j q_j} \< x_i|x_j\> \leq 1$.
where $p_i$ and $q_i$ are probabilities.
Now, the right hand side will not decrease
if we assume $\< x_i|x_j\> = 1$ so we require
$[\sum_i \sqrt{p_i q_i}]^2\leq 1$ which is satisfied by
any probability distribution, which gives the proof 
of  (\ref{pplusfar}) for special $\sigma_{AB}$.

To show the inequality is true in general we find that
\ben
\tr P^+_d \tr_{A'B'}U^{\dagger}\sum_k p_k |\psi_k\>\<\psi_k| U =  \nonumber\\
\sum_k p_k  \tr P^+_d \tr_{A'B'}U^{\dagger}|\psi_k\>\<\psi_k| U.
\een
Thus if (\ref{pplusfar}) holds for $\sigma_{AB}$ derived from pure (product) state, 
by averaging over probabilities, we will have (\ref{pplusfar}) for 
an arbitrary $\sigma_{AB}$ from the set $T$. 

Now by concavity of logarithm, we have for any states $\rho$ and $\sigma$:
\ben
S(\rho||\sigma) = -S(\rho) - \tr(\rho \log \sigma) \geq \nonumber\\
-S(\rho) - \log (\tr \rho \sigma) 
\label{concavlog}
\een
Applying inequality (\ref{pplusfar}) we have that
\be
-\log (\tr \rho \sigma) \geq \log d.
\ee
Now by (\ref{concavlog}) we have that 
\be
S(P^+_d|| \sigma_{AB}) \geq \log d,
\ee
which is a desired bound.}\blacksquare

\subsection*{Acknowledgements}
We would like to thank  Matthias Christandl, Ryszard Horodecki, Andrzej Szepietowski and Andreas Winter for helpful discussion. 
We also thank Geir Ove Myhr for fixing numerous bugs which appeared in  the first version of the manuscript. This work is supported by
the Polish Ministry of Scientific Research and Information Technology under the (solicited) grant No. PBZ-MIN-008/P03/2003,
EU grants RESQ (IST-2001-37559), QUPRODIS (IST-2001-38877) and 
PROSECCO (IST-2001-39227). JO 
acknowledges the Cambridge-MIT Institute. We acknowledge
hospitality of the Isaac Newton Institute for Mathematical Sciences during the QIS programme where the part of this work was done.

\bibliographystyle{IEEEtrans}
\bibliography{rmp12-hugekey}

\end{document}